\documentclass[useAMS,usenatbib]{mnras}
\usepackage[british]{babel}
\usepackage{graphicx}
\usepackage{epstopdf}
\usepackage{color}
\usepackage{amsmath}
\usepackage{amssymb}
\usepackage{array}

\newcommand{\swift}{{\it Swift}}
\newcommand{\swiftbat}{{\it Swift}/BAT}
\newcommand{\swiftxrt}{{\it Swift}/XRT}

\title[2015 outburst of GS 1354--64]{A ``high-hard'' outburst of the black hole X-ray binary GS 1354--64}

\author[K. I. I. Koljonen et al.]
{K.~I.~I.~Koljonen$^{1}$\thanks{email: karri.koljonen@nyu.edu}, D.~M.~Russell$^{1}$, J.~M.~Corral-Santana$^{2}$, M.~Armas Padilla$^{3,4}$, \and T.~Mu\~noz-Darias$^{3,4}$, F.~Lewis$^{5,6}$, M.~Coriat$^{7}$, F.~E.~Bauer$^{2,8,9}$ \\
$^{1}$New York University Abu Dhabi, PO Box 129188, Abu Dhabi, UAE \\
$^{2}$Instituto de Astrof\'isica, Facultad de F\'isica, Pontificia Universidad Cat\'olica de Chile (IA-PUC), Casilla 306, Santiago 22, Chile \\
$^{3}$Instituto de Astrof\'isica de Canarias, 38205 La Laguna, Tenerife, Spain \\
$^{4}$Departamento de astrof\'isica, Univ. de La Laguna, E-38206 La Laguna, Tenerife, Spain \\
$^{5}$Faulkes Telescope Project, School of Physics \& Astronomy, Cardiff University, The Parade, Cardiff, CF24 3AA, Wales \\
$^{6}$Astrophysics Research Institute, Liverpool John Moores University, 146 Brownlow Hill, Liverpool L3 5RF, UK  \\
$^{7}$Institut de Recherche en Astrophysique et Plan\'etologie, 9 Avenue du Colonel Roche, BP 44346, 31028 Toulouse Cedex 4, France \\
$^{8}$Millennium Institute of Astrophysics (MAS), Nuncio Monse\~{n}or S\'{o}tero Sanz 100, Providencia, Santiago de Chile \\
$^{9}$Space Science Institute, 4750 Walnut Street, Suite 205, Boulder, Colorado 80301}

\begin{document}

\pagerange{\pageref{firstpage}--\pageref{lastpage}}
\pubyear{2014}

\maketitle

\label{firstpage}

\begin{abstract} 

We study in detail the evolution of the 2015 outburst of GS 1354--64 (BW Cir) at optical, UV and X-ray wavelengths using Faulkes Telescope South/LCOGT, SMARTS and \swift. The outburst was found to stay in the hard X-ray state, albeit being anomalously luminous with a peak luminosity of L$_{X} >$ 0.15 L$_{Edd}$, which could be the most luminous hard state observed in a black hole X-ray binary. We found that the optical/UV emission is tightly correlated with the X-ray emission, consistent with accretion disc irradiation and/or a jet producing the optical emission. The X-ray spectra can be fitted well with a Comptonisation model, and show softening towards the end of the outburst. In addition, we detect a QPO in the X-ray lightcurves with increasing centroid frequency during the peak and decay periods of the outburst. The long-term optical lightcurves during quiescence show a statistically significant, slow rise of the source brightness over the 7 years prior to the 2015 outburst. This behaviour as well as the outburst evolution at all wavelengths studied can be explained by the disc instability model with irradiation and disc evaporation/condensation.

\end{abstract}


\section{Introduction} \label{introduction}

Transient low-mass X-ray binaries (LMXBs) produce outbursts reminiscent of dwarf novae, but with intervals of typically several years, higher X-ray luminosities usually approaching the Eddington limit, and decays that can last months (e.g. \citealt{chen97}). They are thought to arise from instabilities occurring in the accretion disc, which result in a sudden increase of mass accretion rate onto the compact object (e.g. \citealt{lasota01}). During an outburst, LMXBs often follow a pattern of X-ray spectral changes resembling each other \citep[e.g.][]{maccarone03,belloni10}, where the spectra during the rise of the outburst is dominated by a hard, powerlaw-like spectral component (usually taken to represent the inverse Comptonisation of soft seed photons in a plasma cloud of hot electrons), while the spectra during the peak and initial decay of the outburst is dominated by a soft, blackbody-like spectral component (representing the accretion disc). However, many LMXBs show outbursts that remain in the hard X-ray state for the whole duration of the outburst without changing to the soft X-ray state \citep[][and references therein]{tetarenko15}. 

GS 1354--64 (BW Cir) is a dynamically confirmed black hole LMXB, with a black hole mass $> 7 M_{\odot}$ in a $\sim$ 2.5 day long orbit with a low-mass donor star of type G0--5 III \citep{casares04,casares09}. The distance to the source is likely greater than 25 kpc and less than 61 kpc \citep{casares04,casares09}, but the estimate is heavily dependent on the extinction (lower limit to the distance could be as low as 15 kpc, see \citealt{reynolds11}). At least two previously confirmed outbursts have been observed from GS 1354--64: in 1987 \citep{makino87,kitamoto90} and 1997 \citep{brocksopp01}. The position of GS 1354--64 is also consistent with two earlier outbursts detected from MX 1353--64 \citep{markert77,markert79} and Cen X-2 \citep{francey71}. However, it is not clear if these are connected to GS 1354--64 due to large error boxes in the instruments used. The Cen X--2 outburst in 1967 was reported to be very bright ($\sim$8 Crab; \citealt{brocksopp01}). However, due to the large distance of GS 1354--64, if Cen X-2 was indeed the same source, the outburst was super-Eddington unless the distance to the source is $\sim$15 kpc and the mass of the compact object more than 30$ M_{\odot}$, which seems highly unlikely. The two confirmed outbursts displayed completely different behaviour. In the 1987 outburst the source transitioned to the soft X-ray state with the accretion disc reaching the innermost stable orbit \citep{kitamoto90}. On the other hand, in the 1997 outburst, the source stayed in the hard X-ray state throughout the outburst displaying a pure Comptonisation spectra \citep{revnivtsev00}. The outburst profile of the X-ray lightcurve in the 1997 burst was triangle shaped, and just a single flare was evident, unlike in many other transients that display fast-rise exponential-decay lightcurves or multiple flares and complicated flare profiles. 
 
The optical magnitudes of GS 1354--64 in late May 2015 were reported to be 1.5--2.0 magnitudes brighter than the quiescent values \citep{russell15}, and a new outburst start could be noticed in the Monitor of All-sky X-ray Image (MAXI) and Burst Alert Telescope onboard \swift\/ (\swiftbat) lightcurves approximately 7 days later as reported in \citet{russell15}, but in retrospect the start of the rise in X-rays can be placed approximately 3--4 days later from the first optical detection. The outburst peaked about 1.5 months later \citep{koljonen15} and the decay lasted about three months. 

In this paper, we study the evolution of the 2015 outburst in the optical, ultraviolet (UV) and X-ray frequencies. Multiwavelength observations are important to distinguish different emission scenarios occurring during the outburst. We use the X-ray spectra and lightcurves from pointing and monitoring observations to constrain the X-ray state and to follow the overall evolution to study whether the source exhibits possible X-ray state changes during the outburst. Several emission processes are likely contributing to the optical/UV bands of LMXBs. The usual candidates are thermal emission from the viscously heated accretion disc \citep{shakura73,frank02}, X-ray irradiation and reprocessing of the outer disc \citep{vanparadijs94,dubus99}, or synchrotron emission from the compact jet \citep[e.g. ][]{hynes00,markoff01,russell10,russell13}. Also, other emission mechanisms have been suggested: magnetic loop reconnection (e.g. \citealt{zurita03}), magnetically dominated compact corona (e.g. \citealt{merloni00}), synchrotron emission from a hot inner accretion flow \citep{veledina13} and emission from an advective region (e.g. \citealt{shahbaz03}). In this paper, we use simultaneous optical, UV and X-ray monitoring observations and their correlations to each other, as well as broadband spectral energy distributions (SED), to constrain the optical emission processes. 

The structure of the paper is the following. In Section \ref{observations}, we describe in detail the optical/UV/X-ray observations and how the data were reduced. The outburst characteristics, results from the X-ray spectral and timing analysis, optical/UV/X-ray correlations, broadband SED and long term optical lightcurves are presented in Section \ref{results}. We found that the 2015 outburst, similar to the 1997 outburst, stayed in the hard X-ray state, and that the optical and UV fluxes are tightly correlated with the X-ray flux. In addition, we found a statistically significant, gradual rise in the optical brightness during quiescence. In Section \ref{discussion}, we discuss the implications of these results, and in Section \ref{conclusions} we present our conclusions and discuss the ramifications of our results.        



\section{Observations and data reduction} \label{observations}

\begin{figure}
\begin{center}
\includegraphics[width=0.5\textwidth]{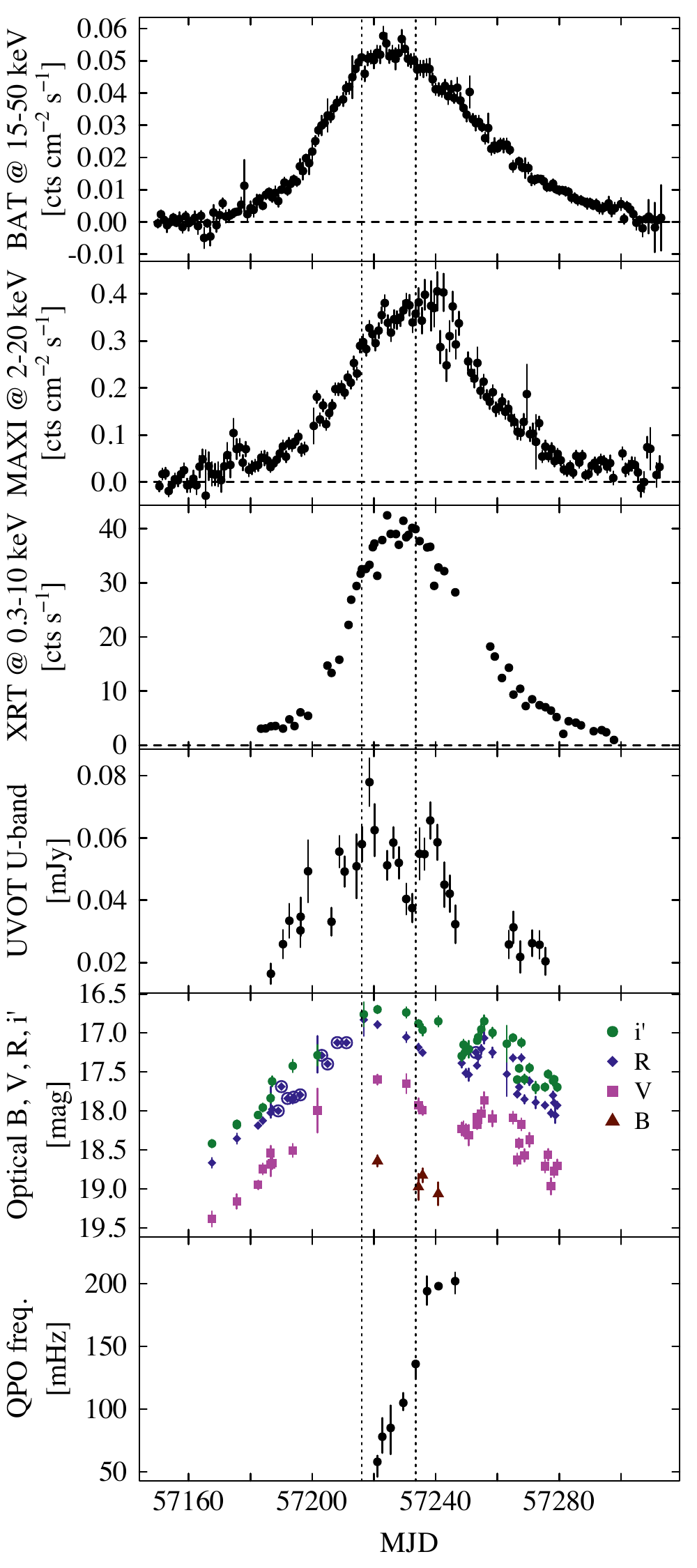}
\end{center}
\vspace{-12pt}
\caption{Monitoring observations of the outburst in X-rays/UV/optical. From top to bottom: \swiftbat\/ 15--50 keV flux, MAXI 2--20 keV flux, \swiftxrt\/ 0.3-10.0 keV count rate, UVOT $U$-band flux, optical magnitudes in $B$- (brown triangles), $V$- (magenta squares), $R$- (dark blue diamonds; SMARTS data are encircled) and $i'$-bands (green circles), and QPO centroid frequency from \swiftxrt\/ timing analysis. The peak of the outburst as determined from the \swiftbat\/ lightcurve is delineated with dotted lines.} \label{plotdata}
\end{figure}

\subsection{Optical/UV observations} \label{optmon}

\subsubsection{Faulkes/LCOGT}

Imaging observations of GS 1354--64 were taken with the 2-m Faulkes Telescope South (FTS) situated at Siding Spring, Australia since 2008 January (MJD 54510), as part of a monitoring campaign of $\sim 40$ LMXBs \citep[see][]{lewis08}. Here we present more than seven years of monitoring data of GS 1354--64 in quiescence, intense monitoring of the 2015 outburst, and monitoring during quiescence post-outburst. During quiescence, observations were typically made once per week when the source was visible, using Bessell $V$, Bessell $R$ and Sloan Digital Sky Survey (SDSS) $i^{\prime}$ filters, with exposure times of 100 sec (from 2008 January 10 to 2008 March 24) or 200 sec (from 2008 March 31 to 2015) per filter on each date. Once the 2015 outburst was detected, we extended our campaign to make use of the suite of 1-m telescopes on the Las Cumbres Observatory Global Telescope Network (LCOGT) and increased the cadence of our monitoring to $\sim 3$ pointings per week for $\sim 100$ days, adding the Bessell $B$-band filter on four dates. From 2015 June 11 to 2015 September 15, data were acquired from the 1-m telescopes situated at Cerro Tololo (Chile), Siding Spring (Australia) and the South African Astronomical Observatory (SAAO), Sutherland (South Africa) as well as the 2-m FTS. Exposure times were generally either 100 sec or 200 sec per filter during outburst. In addition, a sequence of 15 consecutive $i^{\prime}$-band images (100 sec each) were taken with the 1-m at SAAO on 2015 August 1, to assess the short term variability properties of the source in outburst. The source was not visible from the ground from mid-September until December. The pixel scale is 0.30 arcsec pixel$^{-1}$ for the 2-m FTS imaging and 0.47 arcsec pixel$^{-1}$ for the cameras on the 1-m network.

Automatic pipelines de-bias and flat-field the science images using calibration files from the beginning and the end of each night. The source is faint in quiescence, and lies in a crowded region of the Galactic plane, and detections of the source depended on the conditions during each night and the quality of the images. Poor seeing, high airmass and dust extinction (especially in $V$-band) were responsible for the source not being detected on all dates. During the quiescent period 2008--2015, secure detections of GS 1354--64 were achieved from 56 images in $i^{\prime}$-band, 35 images in $R$ and 10 in $V$-band. During the 2015 outburst, 51, 37, 35 and 4 detections were made in $i^{\prime}$, $R$, $V$ and $B$, respectively.

We performed aperture photometry using \small PHOT \normalsize in \small IRAF \normalsize adopting a fixed aperture radius of 6.0 pixels (1.82 arcsec) for the FTS images and 4.5 pixels (2.10 arcsec; different pixel scale) for the 1-m images. These aperture sizes were chosen to ensure almost all of the flux of the source was inside the aperture, while minimising the contamination from nearby sources in this crowded field even under poor seeing conditions. The mean seeing is slightly worse for the 1-m network images, and the aperture size in arcsec is therefore slightly larger. For each image in each filter, the same size aperture was used for photometry on two comparison stars in the field with known magnitudes \citep{brocksopp01,casares09} and used for relative photometry to obtain magnitudes of GS 1354--64. Magnitudes of GS 1354--64 were obtained from 228 usable images in total, including a few observations with larger error bars due to bad conditions. Lists of all the observations for all filters can be found in Tables \ref{fts_i}, \ref{fts_R}, \ref{fts_V} and \ref{fts_B}.    

\subsubsection{SMARTS}

The 1.3m SMARTS data were obtained with A Novel Dual Imaging Camera (ANDICAM) using the Bessell $R$ filter. The data are automatically bias subtracted and flat field calibrated by the SMARTS consortium. The photometry was obtained using two different methods: (i) through point spread function (PSF) fitting using \small DAOPHOT II/ALLSTAR \normalsize \citep{Stetson1987} and (ii) aperture photometry using both \small DAOPHOT II \normalsize and the \small PHOT \normalsize task in \small IRAF\normalsize. In addition, for each aperture photometry, we used two different ways to obtain the final photometry. First, we tried an aperture fixed to 6 pixels (2.21 arcsec) in all images and later we used the optimal aperture for each image following \citet{Naylor1998}. However, the most accurate photometry was obtained using the PSF method. Therefore, we decided to continue the analysis with this one. The final PSF model was obtained using an iterative code to select the best fit (i.e. minimum $\chi^2$) among the six different functions available in \small DAOPHOT II \normalsize (see \citealt{Stetson1987} for further details). To calibrate the field, we selected the five comparisons listed in \citet{casares09} and performed differential photometry. The final magnitude of BW~Cir is then obtained as the weighted mean of the five individual values for each image. The total number of usable images amount to 10, with 9 taken during the outburst rise from MJD 57189 to MJD 57211, and one on the decay on MJD 57253. We do not detect any significant deviation in the magnitudes observed with SMARTS compared to those observed with Faulkes/LCOGT based on the outburst evolution in the optical R-band. However, we do not have strictly simultaneous observations, so direct comparison is not possible. A list of all the observations can be found in Table \ref{smarts}. 

\subsubsection{\swift/UVOT}

In addition, we gathered publicly available pointing observations from the X-ray observatory \swift\/ during the whole outburst. In most of the pointings, observations from the UV/Optical Telescope (UVOT) were available in the $U$-band. We obtained the source fluxes and magnitudes on the level II pipeline processed image files using \textsc{uvotsource}, where we used an aperture of 5 arcsec centered on the source. As GS 1354--64 is located in a dense region of the Galactic plane, we select an annulus from 13.3 arcsec to 18.5 arcsec for the background region that does not include sources. This ensures also that the diffraction features from nearby bright stars do not contaminate the flux estimates. We select only those pointings where the source is detected at 5$\sigma$ level or greater above the background resulting in 31 usable images.     

\subsubsection{Flux determination}

The magnitudes are converted to flux densities using the zero point for each filter (4137 Jy, 3788 Jy, 2948 Jy, 3631 Jy, for $B$-, $V$-, $R$-, and $i^{\prime}$-bands respectively), and dereddened using the \citet{fitzpatrick99} parameterisation and an extinction coefficient of $A_{V} = 2.6\pm0.31$ \citep{casares04,casares09}. This extinction coefficient is measured from the interstellar lines in the optical spectrum. There are no estimates on the Galactic H I column density at the exact location of GS 1354--64. The nearby measurements inside a circle of 1 degree radius vary in a range 5--9$\times 10^{21}$ cm$^{-2}$ with the nearest value (0.2 deg away) being 6.6$\times 10^{21}$ cm$^{-2}$ \citep{kalberla05}. Using the relation between optical extinction and hydrogen column density \citep{guver09}, these correspond to $A_{V} = 2.2-4.3$ with the nearest value corresponding to $A_{V} \sim 3$, which is fairly close to the extinction measured from the interstellar lines. From X-ray studies of the quiescent X-ray state with \textit{Chandra}, the column density has been estimated as 11$\pm$3$\times 10^{21}$ cm$^{-2}$ \citep{reynolds14}. Our mean value of the column density during the outburst is 7.4$\times 10^{21}$ cm$^{-2}$ (Section \ref{xray}), consistent with the Galactic value. In addition, the extinction coefficient cannot be much larger than 3.5, as it would make the optical/UV SED too steep for any physical process ($\alpha>3$) Thus, we use $A_{V} = 2.6$ in the paper.


\subsection{X-ray observations}

We used X-ray monitoring data from MAXI/GSC \citep{matsuoka09} and \swiftbat\/ \citep{krimm13} and obtained the daily 2--20 keV and 15-50 keV fluxes from their web interfaces\footnote{MAXI: http://maxi.riken.jp, \swiftbat: http://swift.gsfc.nasa.gov/results/transients/weak/Ginga1354-645/}, respectively. In addition, we gathered publicly available \swift\/ pointing observations from HEASARC during the whole outburst. The X-Ray Telescope (\swiftxrt) windowed timing (WT) mode data was processed using \textsc{xrtpipeline} in \textsc{heasoft 6.16}, and subsequently the source and background spectrum and response files were extracted using \textsc{xrtproducts}. Exposure maps were generated separately for each pointing. In addition, we extracted 1.8 ms lightcurves to be used in a timing analysis. The spectral and timing analysis was undertaken using \textsc{ISIS} (Interactive Spectral Interpretation System, \citealt{houck00}) with the \textsc{SITAR} timing analysis module. For spectral fitting we selected those spectra between MJD 57206--57272 that have exposures $\gtrsim$ 500 s. We bin the \swiftxrt\/ spectra so that each bin has S/N $\geq$ 20, and we use the channels with energies from 0.8 to 10 keV. We constructed the power density spectrum (PDS) for each pointing using segments of length 32768 bins over the whole lightcurve rejecting the segments with data gaps and averaging over the whole lightcurve. Subsequently, we fit a simple model to the PDS with \textsc{ISIS}, binning the PDS logarithmically with $\delta f/f = 0.1$. The model includes two zero frequency lorentzians, one for modeling the Poisson noise with a deteriorating tail in the higher frequencies because of \swiftxrt\/ instrument read-out method, and the other for modeling flat top noise in the lower frequencies. In addition, a number of lorentzians are added to model the QPOs and their harmonics when necessary. 

The Burst Alert Telescope (\swiftbat) data were processed using \textsc{batbinevt v1.43} on the survey mode data files (DPH) to first produce sky image files (DPI), which are used to make a masked weighting map for the position of GS 1354--64 with \textsc{batmaskwtimg v1.19}. \textsc{batbinevt v1.43} is subsequently used to process the DPH files to PHA files using the masked weighting map. A systematic error vector is applied to the spectra to account for residuals in the response matrix using \textsc{batphasyserr v1.4}, and subsequently responses were created with \textsc{batdrmgen v3.4}. If more than one spectrum is obtained during a single pointing these were summed together. For \swiftbat\/, we use the channels with energies from 15 to 100 keV.

\section{Results} \label{results}

\subsection{Outburst characteristics}

\begin{figure}
\begin{center}
\includegraphics[width=0.5\textwidth]{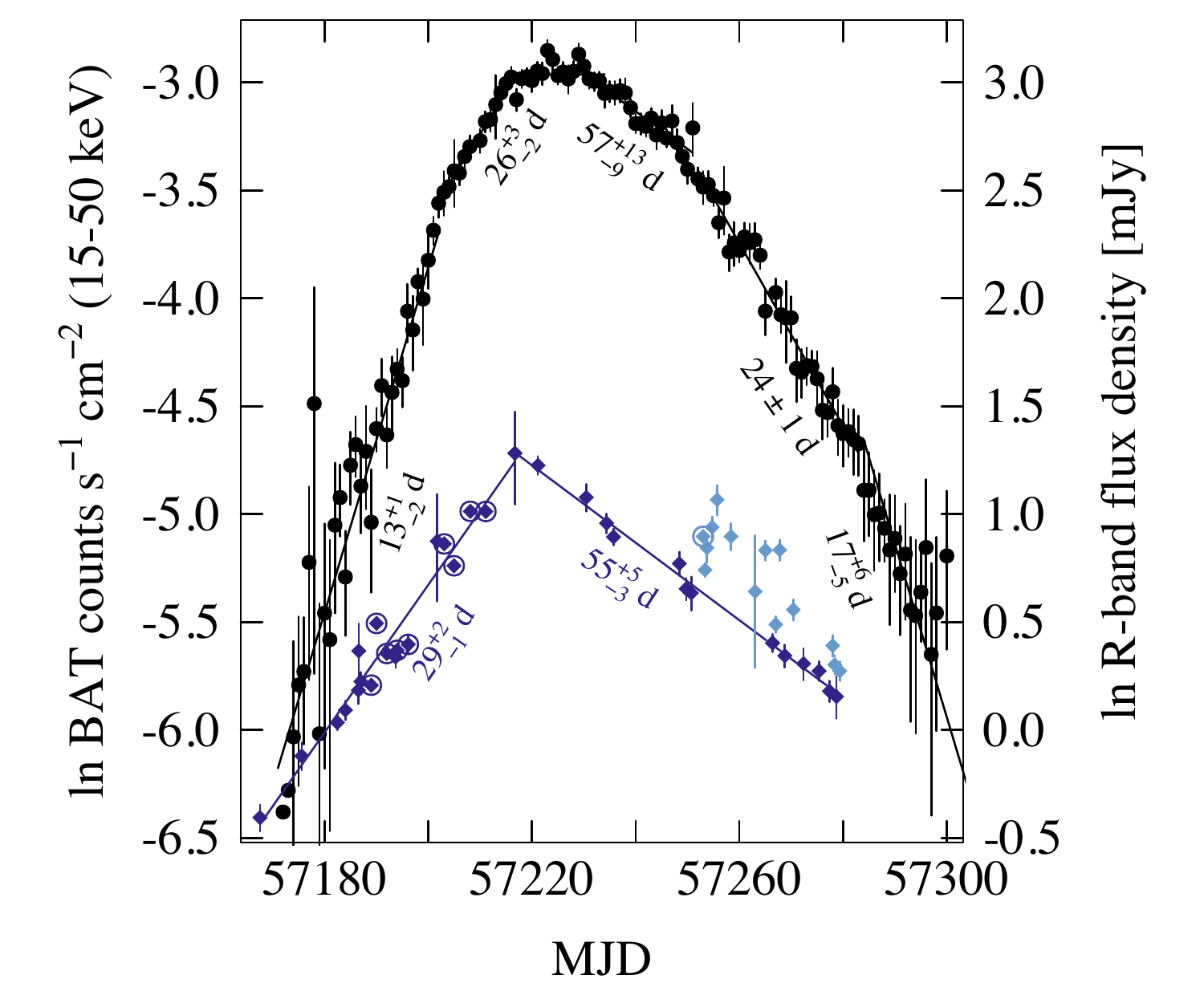}
\end{center}
\vspace{-12pt}
\caption{\swiftbat\/ (black circles) and dereddened optical $R$-band (blue diamonds; SMARTS data are encircled) lightcurves of the outburst overplotted. In addition, the e-folding times for the exponential rise and decay are marked for both outburst profiles. The light-blue diamonds are excluded in the calculation of the exponential decay in the $R$-band.} \label{plotbat}
\end{figure}

The lightcurves of the 2015 outburst are plotted in Fig. \ref{plotdata} for X-rays, UV and optical bands. The outburst morphology in hard X-rays is triangular in the nomenclature of \citet{chen97} with a flat top. The outburst rise as observed by \swiftbat\/ lasts 46 days from MJD 57171 to MJD 57217 and consists of two parts with the flux exponentially increasing with a different e-folding time (see Fig. \ref{plotbat}): 13$^{+1}_{-2}$ and 26$^{+3}_{-2}$ days, respectively, with the change taking place on MJD 57202. During the outburst peak that lasts 17 days from MJD 57217 to MJD 57234, the hard X-ray flux is flat with two small flares. The mean \swiftbat\/ flux at the peak is 0.052 $\pm$ 0.003 cts/cm$^2$/s. After the peak, the hard X-ray flux decays with an e-folding time of 57$^{+13}_{-9}$ days until MJD 57251 where it changes to 24$\pm$1 days and again to 17$^{+6}_{-5}$ days from MJD 57284 onwards. The soft X-ray lightcurve is almost identical to the hard X-ray with the outburst peak slightly lagging because of the softening of the X-ray spectra (see Section \ref{xray}). The mean MAXI flux at the peak is 0.34 $\pm$ 0.03 cts/cm$^2$/s. The two X-ray flux regimes are strongly correlated with a Spearman rank correlation coefficient of 0.89$\pm0.01$. The outburst, rise and peak durations correspond well to the previous outburst in 1997 \citep{brocksopp01}. However, the 2015 outburst is 3--4 times brighter in the soft X-ray band (2-12 keV) than the outburst in 1997. There are no detectable drops in the \swiftbat\/ lightcurve, thus the source does not undergo any state transitions during the outburst.   

\begin{figure*}
\begin{center}
\includegraphics[width=1.0\textwidth]{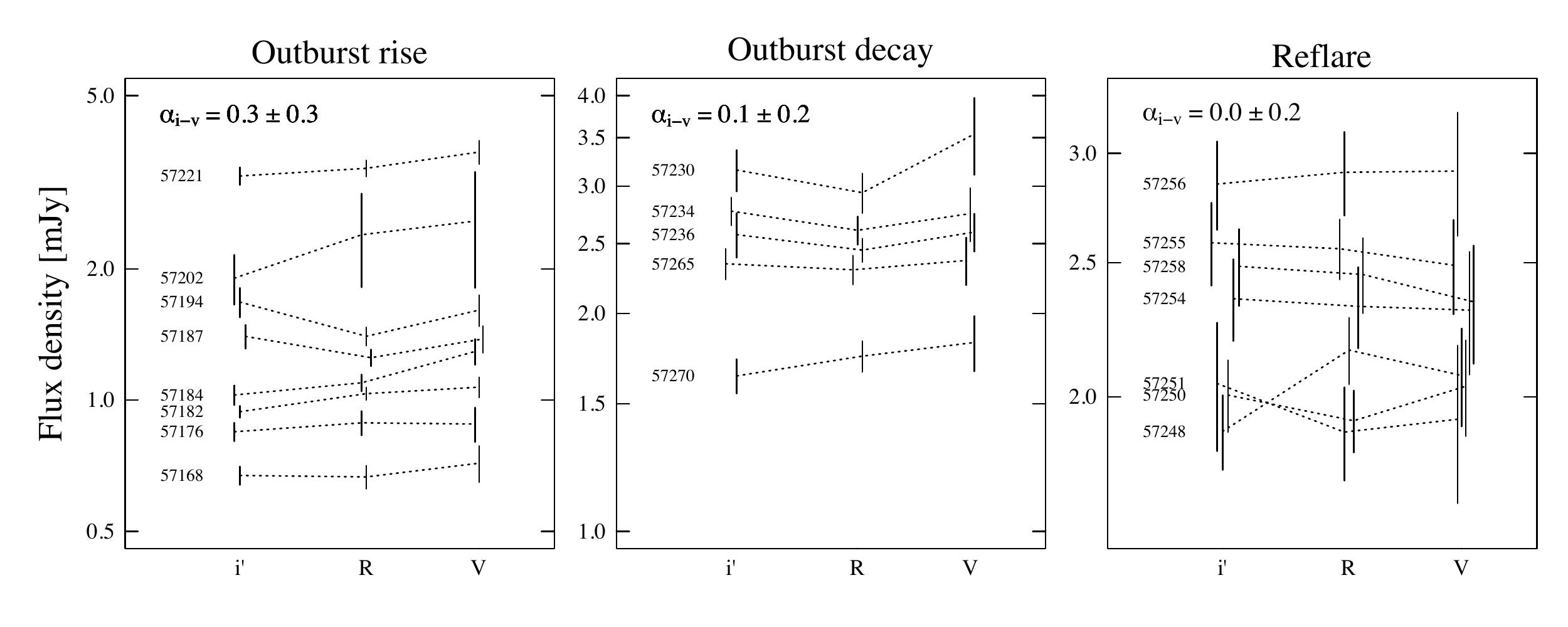}
\end{center}
\vspace{-12pt}
\caption{Dereddened optical SEDs for the outburst rise (left), outburst decay (middle), and reflare(s) during decay (right). The subsequent spectra have been shifted vertically for clarity. The average spectral index between i'- and V-band is marked in the upper-left corner of each panel.} \label{optsed}
\end{figure*}

The optical lightcurves present similar morphology to the X-rays, and rise and peak in unison with the X-rays indicating correlated behaviour (see Section \ref{optx}). The first detection above the quiescent level in optical wavelengths on MJD 57167 is $\sim$1.5 magnitudes brighter than the mean quiescent values, and it was observed 3--4 days before a detectable rise in the \swiftbat\/ and MAXI lightcurves. The optical lightcurve exhibits similar exponential flux increases as the X-rays, and interestingly the coefficients are closely similar to the X-ray ones near the outburst peak (see Fig. \ref{plotbat}) with the e-folding time being 29$^{+2}_{-1}$ days during the rise, and 55$^{+5}_{-3}$ days during the decay (we disregard times when the optical is exhibiting reflares). The outburst peak in the optical is sharper compared to the X-ray, except in the $i'$-band which shows a flatter profile. The optical magnitudes at the outburst peak are $m_{V} = 17.60\pm0.07$ mag, $m_{R} = 16.89\pm0.05$ mag, and $m_{i'} = 16.70\pm0.05$ mag. These values are slightly lower when compared to the ones observed in previous outbursts ($m_{V} \leq$ 16.9 in the 1987 outburst, and $m_{V} \leq$ 17.3 in the 1997 outburst; \citealt{brocksopp01}) In Fig. \ref{optsed} we plot the optical SEDs in the rise and decay parts of the outburst. Here, the quasi-simultaneous magnitudes (exposure time and readout time in between) have been converted to fluxes and dereddened as described in Section \ref{optmon}. All the fluxes are highly correlated and the SEDs are flat with $\alpha=0.15\pm0.30$ ($S_{\nu}=\nu^{\alpha}$; or more inverted if using a higher extinction value) during the outburst rise, decay and the optical reflare. During the higher cadence 0.5 hour SAAO observation on MJD 57235, the optical fractional rms variability was measured in $i'$-band to be $<$ 7\%. This is lower than the optical rms of GX 339--4 in the hard state at a similar time resolution (typically $\geq$ 10\%; \citealt{cadollebel11}).

\begin{figure}
\begin{center}
\includegraphics[width=0.5\textwidth]{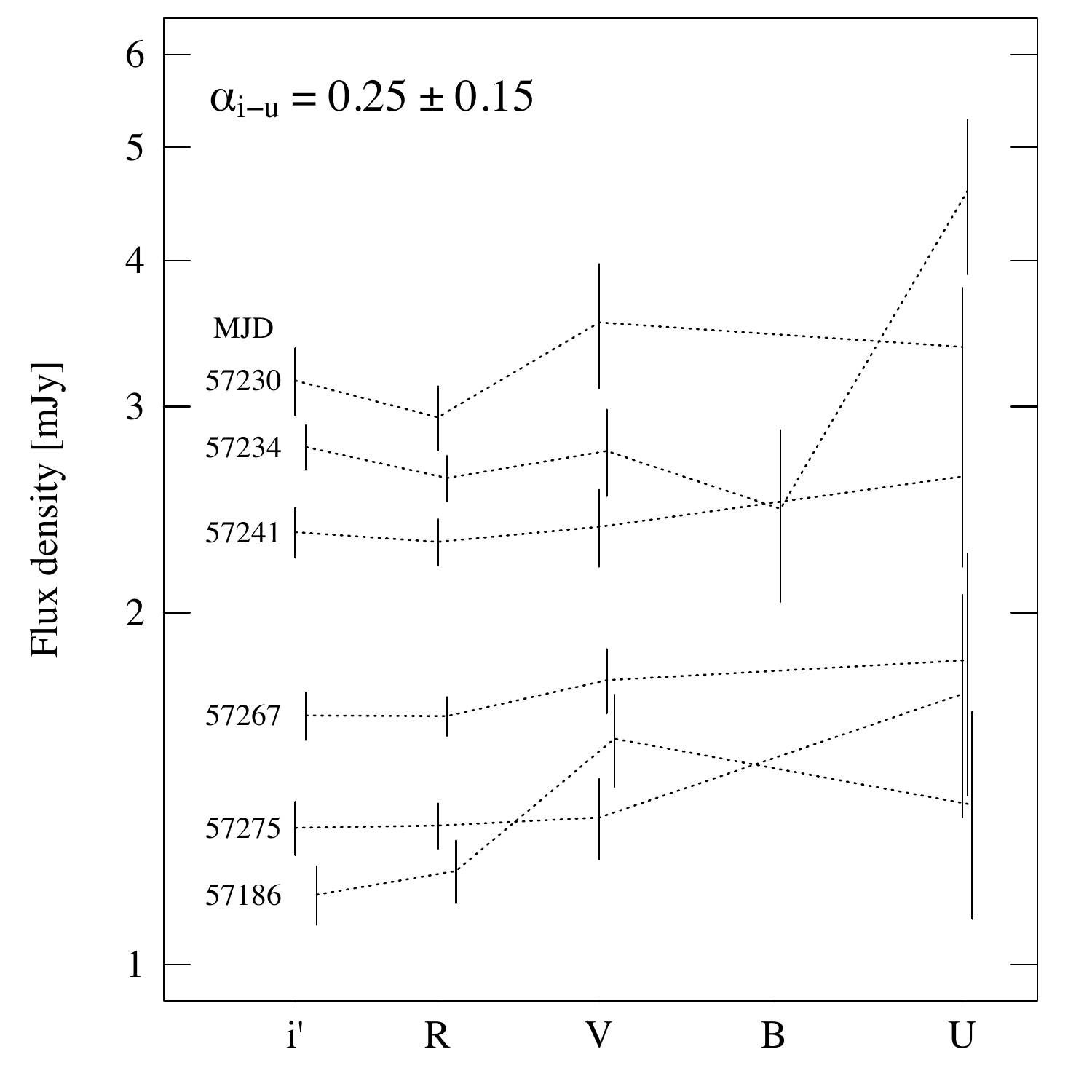}
\end{center}
\vspace{-12pt}
\caption{Dereddened optical/UV SEDs of the outburst when simultaneous (within a day) observations were available. The subsequent spectra have been shifted vertically for clarity. The average spectral index between i'- and U-band is marked in the upper-left corner.} \label{optsed_u}
\end{figure}

The UV lightcurve also shows a similar outburst profile as the optical and X-ray. The UV lightcurve of the outburst rise and decay exhibit similar exponential behaviour as the optical lightcurves with an e-folding time of $\sim$30 days and $\sim$56 days during the rise and decay, respectively. The mean flux density during the hard X-ray and optical peak is $F_{U} \sim$ 0.06 mJy, but the UV exhibits moderate flaring in the peak with the flux changing from 0.04 mJy to 0.08 mJy. During the decay there is a strong flare around MJD 57240, but unfortunately there is a gap in optical monitoring during that time apart from a single observation in $i'$- and $B$-band. Similarly, the UV lightcurve has a gap during the optical reflare around MJD 57255, and thus it is difficult to ascertain whether the optical reflare(s) is visible in the $U$-band. Fig. \ref{optsed_u} shows the optical/UV SEDs when simultaneous (within a day) observations were available. The SEDs are flat up to the $U$-band, with the exception of excess in the $V$-band during MJD 57187 and in the $U$-band during MJD 57234. 

\subsection{X-ray spectral and timing analysis} \label{xray}

All \swiftxrt\/ spectra can be fitted with an absorbed powerlaw model ($\Gamma \sim$ 1.2--1.7) excluding a soft excess that seems to originate below 0.8 keV and a collection of emission/absorption lines that are present in some of the spectra, with the most prominent emission line at 2.3 keV (sulphur K$\alpha$, mean equivalent width 32 eV), weaker lines around 1.7--1.9 keV probably arising from silicon, and 6.4 keV iron K$\alpha$ line. Adding the \swiftbat\/ spectrum makes the power law fit worse and a better fit is obtained with a Comptonisation model (\textsc{comptt}; \citealt{titarchuk94}). Many models were tried in addition to \textsc{comptt}, which return similar results. Thus, we fit the individual \swift\/ pointings with a model \textsc{constant $\times$ tbabs $\times$ comptt}, where the constant is allowed for the different calibrations of \swiftxrt\/ and \swiftbat\/. The mean column density is $N_{H} = 7.4 \times 10^{21} \mathrm{cm}^{-2}$ when the absorption is left free in the fits, which is very close to the value obtained from the Galactic H I survey in the field as discussed in Section \ref{optmon}. Thus, we fix the absorption to this value. In addition, the electron temperature ($kT_{e}$) and the optical depth ($\tau$) are degenerate in the fits, thus we fix the $kT_{e}$ to 13 keV (mean value when left free), and fit just the optical depth, and present these as the Compton-y parameter: y = $(4kT_{e}/511) \times \mathrm{max}(\tau,\tau^{2})$. During the outburst rise, the error on the lower bound of the disk temperature was found to be pegged to the minimum value, thus in these cases we fixed the temperature to the upper bound, and proceeded to fit the spectrum again.

\begin{figure}
\begin{center}
\includegraphics[width=0.5\textwidth]{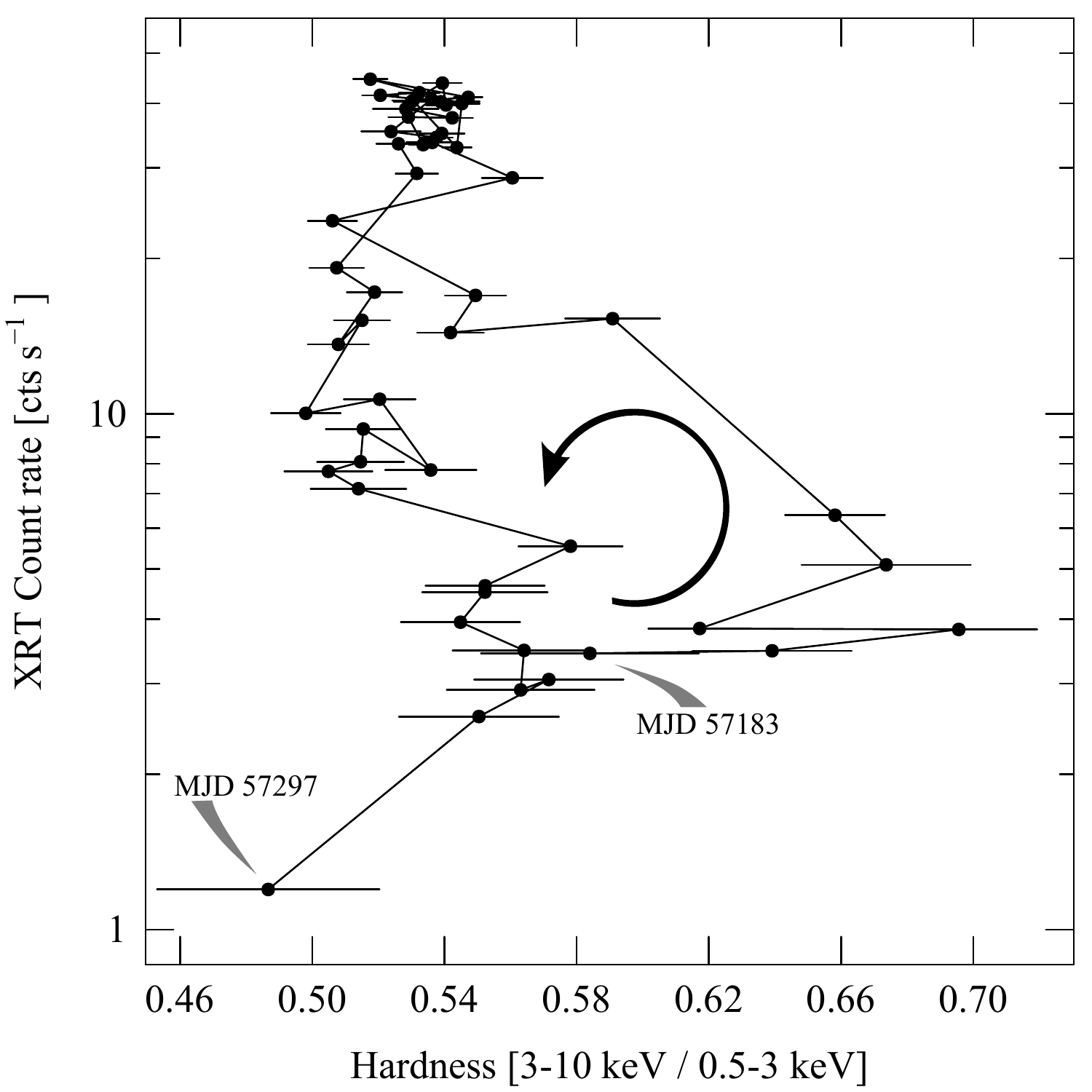}
\end{center}
\vspace{-12pt}
\caption{The hardness-intensity diagram of the outburst using \swiftxrt\/ pointings. The observation date for the first and last pointing are marked in the plot, as well as the direction of the evolution of the hardness ratio.} \label{hr}
\end{figure}

\begin{figure}
\begin{center}
\includegraphics[width=0.5\textwidth]{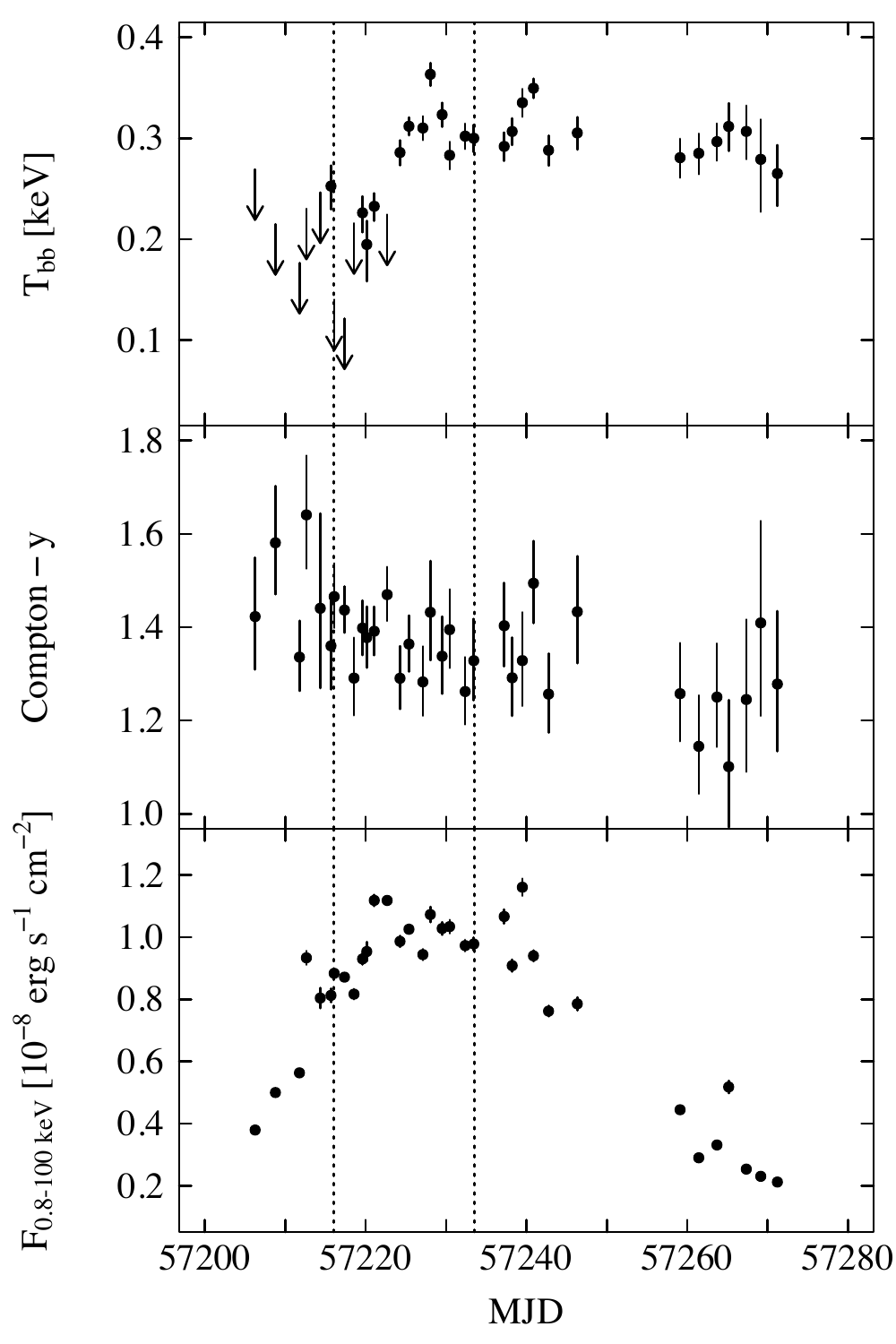}
\end{center}
\vspace{-12pt}
\caption{Model parameter evolution throughout the outburst. From top to bottom: the temperature of the seed black body photons, Compton-y parameter, and the model flux in the energy range 0.8--100 keV. The dotted lines delineate the outburst peak as shown in Fig. \ref{plotdata}.} \label{plotparams}
\end{figure}

The Compton-y changes from $\sim$1.6 to $\sim$1.2 indicating spectral softening throughout the outburst. The spectral softening is evident also in the \swiftxrt\/ hardness-intensity diagram (Fig. \ref{hr}), where the outburst rise shows a harder spectra compared to the outburst decay. As the outburst fades the hardness ratio returns to values observed at the start of the outburst forming a loop in the diagram. However, as the spectral fits from \swift\/ were all fitted well with a Comptonisation model and the disc component was not needed, the outburst is hard: the source remained in the hard X-ray state throughout its 2015 outburst, similar to the one in 1997. The mean unabsorbed flux in the X-ray band 0.8--100 keV is 1.03$\times 10^{-8}$ erg s$^{-1}$ cm$^{-2}$ during MJD 57221--57227, which corresponds to a luminosity of 7.7$\times 10^{38}$ erg s$^{-1}$ at 25 kpc. We fix the \swiftbat\/ spectra to the level of the \swiftxrt\/ spectra in the flux calculation. The parameters and total fluxes of the fits are presented in Table \ref{models} and plotted in Fig. \ref{plotparams}. We would like to note that in a few spectra, a satisfactory fit could not be achieved because of prominent emission lines present in the \swiftxrt\/ spectra. In these cases gaussian profiles were added to the model at the location of the emission lines around 1.7--1.9 keV, 2.3 keV, and/or 6.4 keV. 

\begin{figure}
\begin{center}
\includegraphics[width=0.5\textwidth]{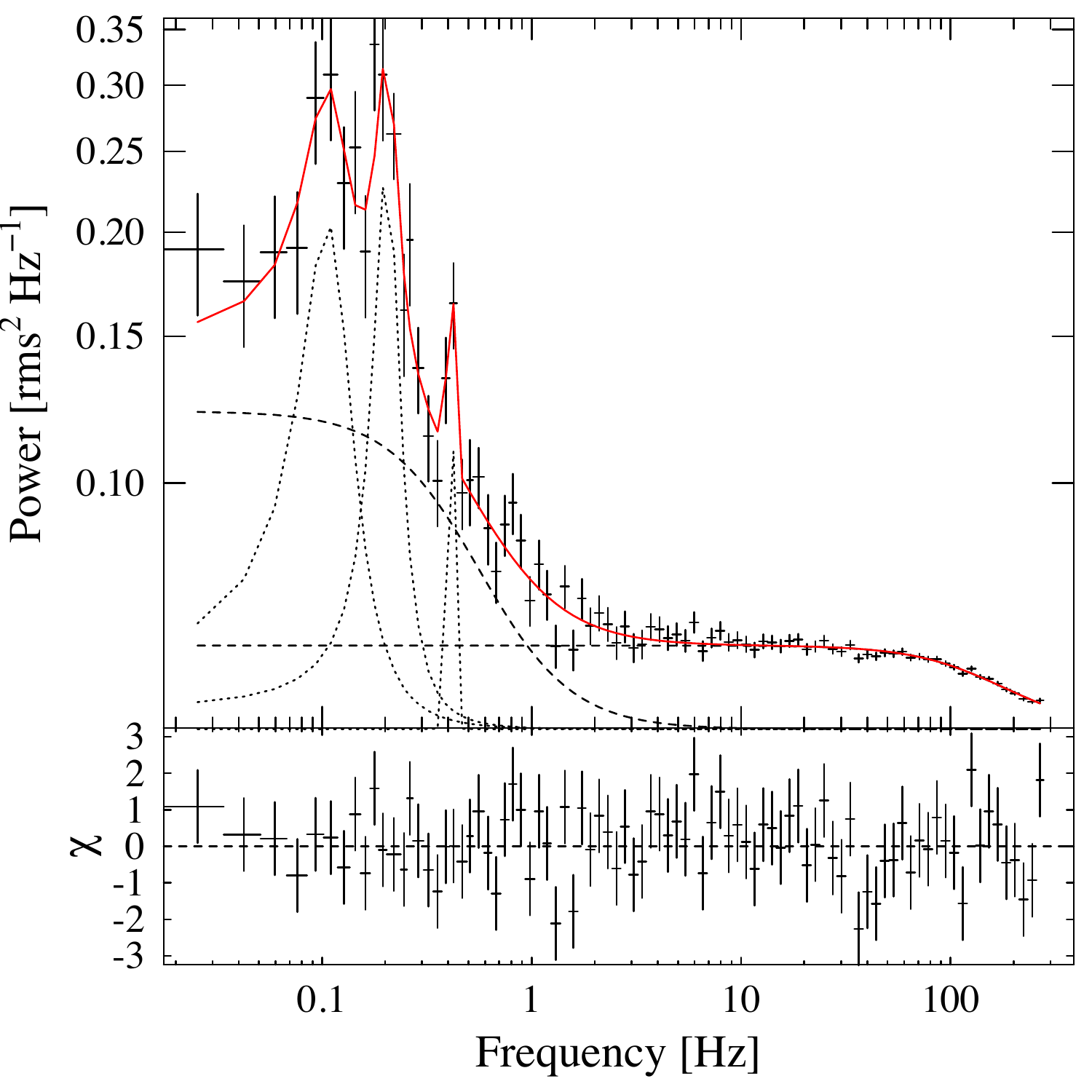}
\end{center}
\vspace{-12pt}
\caption{Power density spectrum (upper panel) from \swiftxrt\/ lightcurve of pointing 00033811041 showing a QPO at 0.2 Hz with a lower and higher harmonic (dotted lines). Red, solid line shows the best fit model, dashed lines the noise components, and the model residuals are shown in the lower panel.} \label{plotqpo}
\end{figure}

In some of the \swiftxrt\/ power density spectra, we can identify a low-frequency QPO (in some cases with harmonics, see Fig. \ref{plotqpo}) that changes in frequency as the outburst evolves. The evolution of the QPO is similar to the outburst in 1997, emerging when the outburst reaches the peak/top part of the lightcurve, increasing in frequency at the top, and settling to an approximately constant value during the decay. In the 1997 outburst this value corresponded to $\sim$ 40 mHz, while in the 2015 outburst the value is five times higher: $\sim$ 200 mHz (Fig. \ref{plotdata}). The disc temperature is also lower ($<$0.3 keV) during the outburst rise and settles to a constant value of $\sim$ 0.3 keV in the outburst top/decay with two flares approximately on MJD 57230 and MJD 57240. Interestingly, these flares occur roughly at the same time as the UV lightcurve exhibits flares and thus indicates that the accretion disc, as a source of the seed photons, is also responsible at least in part for the UV emission.  

\begin{table*} \centering
\caption{X-ray spectral fits. In the model, the interstellar absorption, N$_{H}$, is fixed to 0.74$\times 10^{22}$ cm$^{2}$. The errors are quoted at 90\% confidence level. The columns are: (1) the identification number of the pointing, (2) start time of the pointing in MJD, (3) \swiftxrt\/ and (4) \swiftbat\/ exposure in seconds, (5) the temperature of the seed black body photons, (6) Compton-y parameter, (7) model flux in a range 0.8--100 keV, (8) the reduced chi-squared value, and (9) the centroid frequency of quasi-periodic oscillation if present.} 
\label{models}
\begin{tabular}{cccccccccc} \hline
(1) & (2) & (3) & (4) & (5) & (6) & (7) & (8) & (9) \\
ObsID$^{a}$ & Time [MJD] &  XRT exp. [s] & BAT exp. [s] & kT$_{bb}$ [keV] & y & F$_{X}^{b}$ & $\chi^{2}_{red}$/d.o.f & $f_{\mathrm{QPO}}$ [Hz] \\\hline
11 & 57206.2 & 945 & 971 & $<$0.27 & 1.42$^{+0.13}_{-0.11}$ & 0.38$\pm$0.01 & 1.22/63 & -- \\   
12 & 57208.8 & 964 & 990 & $<$0.21 & 1.58$^{+0.12}_{-0.11}$ & 0.50$\pm$0.01 & 1.08/69 & -- \\ 
14 & 57211.8 & 981 & 956 & $<$0.18 & 1.34$^{+0.07}_{-0.08}$ & 0.56$\pm$0.01 & 1.05/81 & -- \\  
15 & 57212.6 & 591 & 609 & $<$0.23 & 1.64$^{+0.13}_{-0.11}$ & 0.93$\pm$0.02 & 1.04/72  & -- \\ 
18 & 57215.7 & 650 & 667 & 0.25$^{+0.02}_{-0.02}$ & 1.36$^{+0.10}_{-0.09}$ & 0.81$\pm$0.02 & 0.78/81 & -- \\ 
19 & 57216.1 & 827 & 836 & $<$0.14 & 1.47$^{+0.07}_{-0.07}$ & 0.88$\pm$0.02 & 1.26/96 & -- \\ 
20 & 57217.4 & 1582 & 1600 & $<$0.12 & 1.44$^{+0.05}_{-0.05}$ & 0.87$\pm$0.01 & 1.32/146 & -- \\  
21 & 57218.5 & 449 & 465 & $<$0.22 & 1.29$^{+0.09}_{-0.08}$ & 0.82$\pm$0.02 & 0.98/70 & -- \\ 
22 & 57219.6 & 1421 & 1437 & 0.23$^{+0.01}_{-0.02}$ & 1.40$^{+0.06}_{-0.06}$ & 0.93$\pm$0.02 & 1.23/145 & -- \\   
23 & 57220.1 & 854 & 870 & 0.19$^{+0.03}_{-0.03}$ & 1.38$^{+0.06}_{-0.07}$ & 0.95$\pm$0.03 & 0.98/103 & -- \\ 
24 & 57221.1 & 2086 & 2110 & 0.23$^{+0.02}_{-0.01}$ & 1.39$^{+0.05}_{-0.05}$ & 1.12$\pm$0.02 & 1.45/169 & 0.058$^{+0.005}_{-0.012}$ \\ 
25 & 57222.7 & 1175 & 1124 & $<$0.22 & 1.47$^{+0.06}_{-0.06}$ & 1.12$\pm$0.01 & 1.73/123 & 0.078$^{+0.015}_{-0.013}$ \\ 
26 & 57224.3 & 1075 & 1090 & 0.29$^{+0.01}_{-0.02}$ & 1.29$^{+0.07}_{-0.07}$ & 0.99$\pm$0.02 & 1.40/128 & -- \\ 
27 & 57225.4 & 1828 & 1834 & 0.31$^{+0.01}_{-0.01}$ & 1.36$^{+0.07}_{-0.06}$ & 1.03$\pm$0.01 & 1.03/177 & 0.085$^{+0.018}_{-0.021}$\\ 
28 & 57227.1 & 1085 & 1098 & 0.31$^{+0.01}_{-0.01}$ & 1.28$^{+0.08}_{-0.07}$ & 0.94$\pm$0.02 & 1.39/122 & -- \\ 
29 & 57228.0$^{c}$ & 1145 & 1158 & 0.36$^{+0.01}_{-0.01}$ & 1.43$^{+0.11}_{-0.10}$ & 1.07$\pm$0.02 & 1.63/122 & -- \\  
30 & 57229.5 & 954 & 981 & 0.32$^{+0.02}_{-0.01}$ & 1.34$^{+0.08}_{-0.08}$ & 1.03$\pm$0.02 & 1.13/117 & 0.105$^{+0.008}_{-0.006}$ \\ 
31 & 57230.4 & 897 & 921 & 0.28$^{+0.02}_{-0.01}$ & 1.39$^{+0.09}_{-0.08}$ & 1.03$\pm$0.02 & 1.16/112 & 0.105$^{+0.008}_{-0.006}$ \\ 
33 & 57232.4 & 998 & 1024 & 0.30$^{+0.01}_{-0.01}$ & 1.26$^{+0.08}_{-0.07}$ & 0.97$\pm$0.02 & 1.39/120 & -- \\ 
34 & 57233.5 & 831 & 860 & 0.30$^{+0.01}_{-0.01}$ & 1.33$^{+0.09}_{-0.09}$ & 0.98$\pm$0.02 & 1.28/106 & 0.136$^{+0.000}_{-0.012}$ \\  
37 & 57237.2 & 887 & 911 & 0.29$^{+0.02}_{-0.01}$ & 1.40$^{+0.10}_{-0.08}$ & 1.07$\pm$0.02 & 1.55/107 & 0.194$^{+0.012}_{-0.011}$\\  
38 & 57238.2 & 910 & 923 & 0.31$^{+0.01}_{-0.02}$ & 1.29$^{+0.09}_{-0.08}$ & 0.91$\pm$0.02 & 1.24/109 & -- \\ 
39 & 57239.5 & 972 & 951 & 0.34$^{+0.01}_{-0.02}$ & 1.33$^{+0.10}_{-0.10}$ & 1.16$\pm$0.03 & 1.50/98 & -- \\ 
41 & 57240.9$^{c}$ & 2152 & 2177 & 0.35$^{+0.01}_{-0.01}$ & 1.49$^{+0.09}_{-0.09}$ & 0.94$\pm$0.02 & 1.63/168 & 0.198$^{+0.001}_{-0.001}$ \\ 
42 & 57242.8 & 850 & 879 & 0.29$^{+0.01}_{-0.02}$ & 1.26$^{+0.08}_{-0.09}$ & 0.76$\pm$0.02 & 1.36/95 & -- \\ 
44 & 57246.3$^{c}$ & 1058 & 1082 & 0.31$^{+0.02}_{-0.02}$ & 1.43$^{+0.12}_{-0.11}$ & 0.79$\pm$0.02 & 1.27/96 & 0.202$^{+0.007}_{-0.010}$ \\
46 & 57259.1 & 1045 & 1062 & 0.28$^{+0.02}_{-0.02}$ & 1.26$^{+0.11}_{-0.10}$ & 0.44$\pm$0.01 & 1.29/74 & -- \\
47 & 57261.4 & 1104 & 1133 & 0.29$^{+0.01}_{-0.03}$ & 1.14$^{+0.11}_{-0.10}$ & 0.29$\pm$0.01 & 1.38/65 & -- \\
48 & 57263.7 & 1187 & 1191 & 0.30$^{+0.01}_{-0.02}$ & 1.25$^{+0.12}_{-0.11}$ & 0.33$\pm$0.01 & 1.31/72 & -- \\
49 & 57265.2 & 1077 & 1099 & 0.31$^{+0.02}_{-0.02}$ & 1.10$^{+0.14}_{-0.13}$ & 0.52$\pm$0.02 & 1.10/57 & -- \\
50 & 57267.4 & 1004 & 1015 & 0.31$^{+0.02}_{-0.03}$ & 1.25$^{+0.17}_{-0.16}$ & 0.25$\pm$0.01 & 0.94/57 & -- \\
51 & 57269.2 & 917 & 937 & 0.28$^{+0.04}_{-0.05}$ & 1.41$^{+0.22}_{-0.20}$ & 0.23$\pm$0.01 & 1.34/49 & -- \\
52 & 57271.2 & 1091 & 1110 & 0.27$^{+0.02}_{-0.04}$ & 1.28$^{+0.15}_{-0.15}$ & 0.21$\pm$0.01 & 1.10/55 & -- \\
\hline
\end{tabular}
\begin{list}{}{}
\item[$^{a}$] 000338110** \\
\item[$^{b}$] Units are in 10$^{-8}$ erg s$^{-1}$ cm$^{-2}$. \\
\item[$^{c}$] Gaussian profiles added to the model at 1.7--1.9 keV, 2.3 keV and/or 6.4 keV. \\
\end{list}
\end{table*}

\subsection{Optical/UV/X-ray correlation} \label{optx}

\begin{figure*}
\begin{center}
\includegraphics[width=1.0\textwidth]{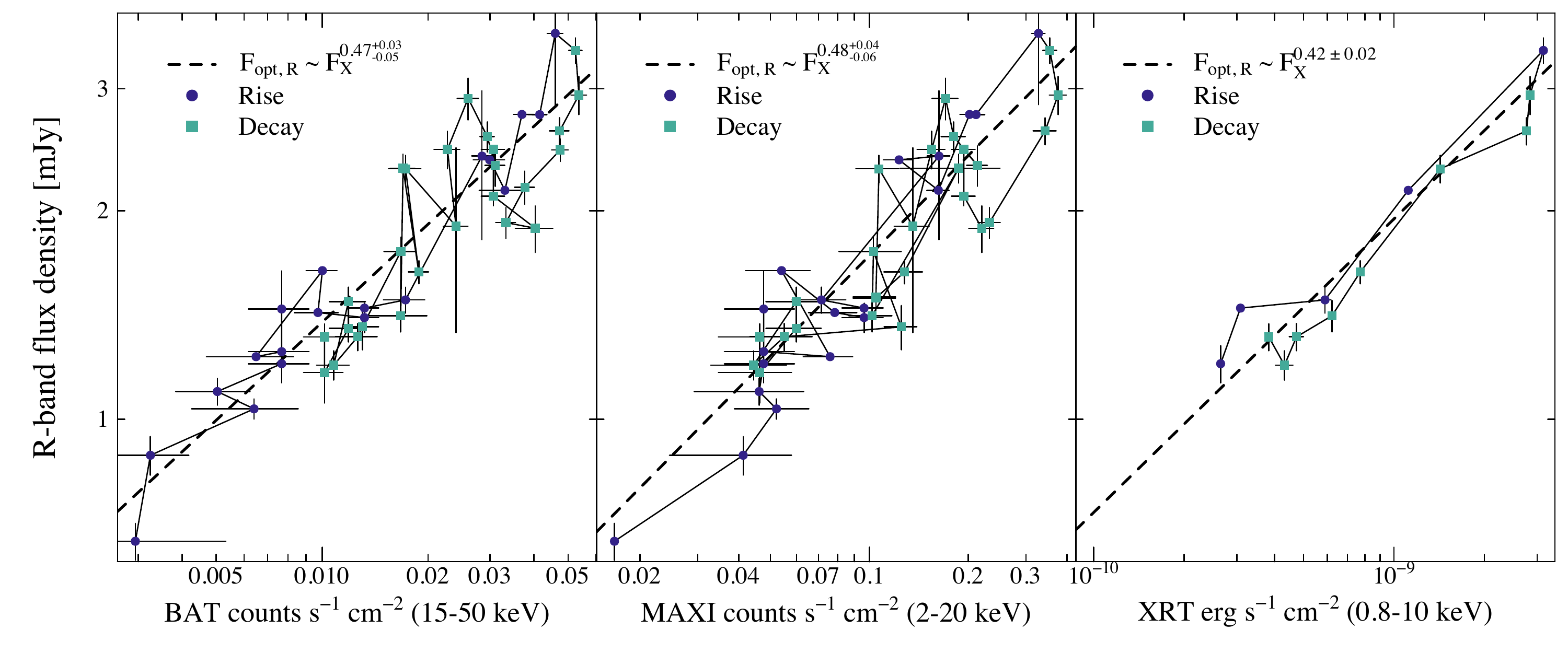}
\end{center}
\vspace{-12pt}
\caption{Optical/X-ray correlation. \textit{Left:} Optical $R$-band flux density as a function of the \swiftbat\/ flux density. The dashed line shows the best least squares fit to the data, with the resulting slope marked in the legend on the upper left. The rise and decay parts of the outburst have been marked with different colours and symbols (blue circles for the rise, and turquoise squares for the decay). \textit{Middle:} Same as in the left panel but the optical $R$-band flux density is plotted as a function of MAXI flux density. \textit{Right:} Same as in the left panel but the optical $R$-band flux density is plotted as a function of the 0.8--10 keV X-ray flux. Due to the requirement that the optical and X-ray observations should be within a day, the number of simultaneous \swift\/ pointings with the optical is lower as compared to the monitoring data.} \label{optXcorr}
\end{figure*}

\begin{figure}
\begin{center}
\includegraphics[width=0.5\textwidth]{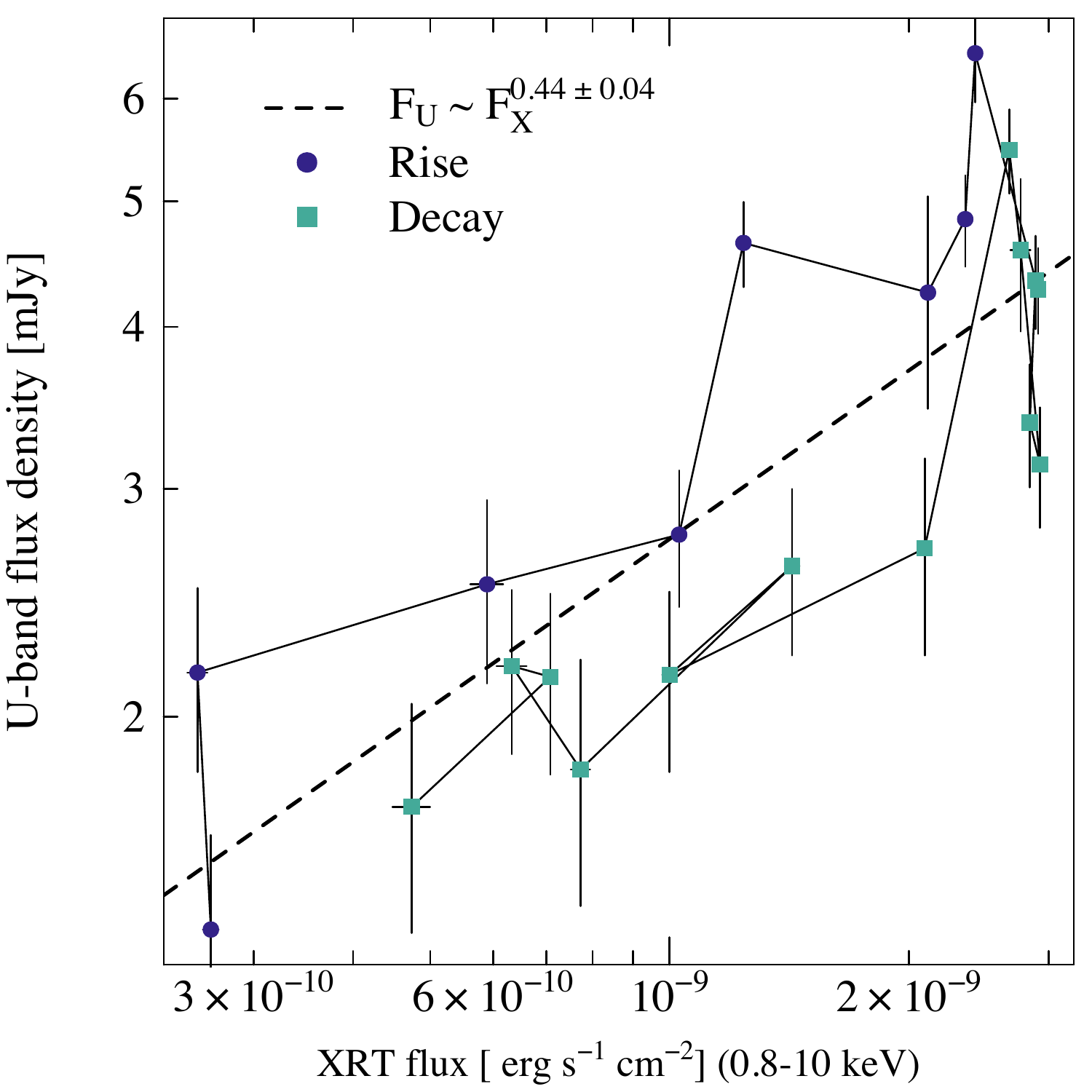}
\end{center}
\vspace{-12pt}
\caption{UV/X-ray correlation. Same as in Fig. \ref{optXcorr}, but $U$-band flux density as a function of the 0.8--10 keV X-ray flux.} \label{UVcorr}
\end{figure}

In Fig. \ref{optXcorr} we show the dereddened optical $R$-band flux density as a function of the X-ray fluxes from MAXI, \swiftbat\/ and \swiftxrt. The X-ray fluxes are restricted to be within a day of the $R$-band observations. For the outburst rise there is a clear correlation between the $R$-band and X-ray fluxes with the relation $F_{opt} \sim F_{X}^{0.4-0.5}$. At the peak of the outburst the $R$-band flux density drops slightly faster than the X-ray fluxes, however returning to the same relation during the optical flare and continuing on it as the outburst decays. During a brief period, as the optical reflare rises in magnitude, the $R$-band flux density is anticorrelated with the X-ray fluxes. The optical/X-ray correlations for $i'$- and $V$-bands are very similar (see Fig. \ref{optixcorr} for $i'$-band including one data point in the quiescent state).

Similar to the optical flux densities, Fig. \ref{UVcorr} shows the dereddened $U$-band flux density as a function of the X-ray fluxes, where the X-ray fluxes are restricted to be within a day of the $U$-band observations. As above, the $U$-band show a clear correlation with the X-ray fluxes with the same relation $F_{U} \sim F_{X}^{0.4-0.5}$ as the optical. Due to the rather large errors and strong flaring of $U$-band at the peak and decay of the outburst, it is difficult to distinguish whether the $U$-band data exhibit a similar drop as the optical flux densities after the outburst peak. Instead, it is possible that the correlations exhibit two tracks, where the outburst rise is brighter in the $U$-band compared to the outburst decay, at a given X-ray flux.  

\subsection{Multiwavelength spectral energy distribution}

\begin{figure}
\begin{center}
\includegraphics[width=0.5\textwidth]{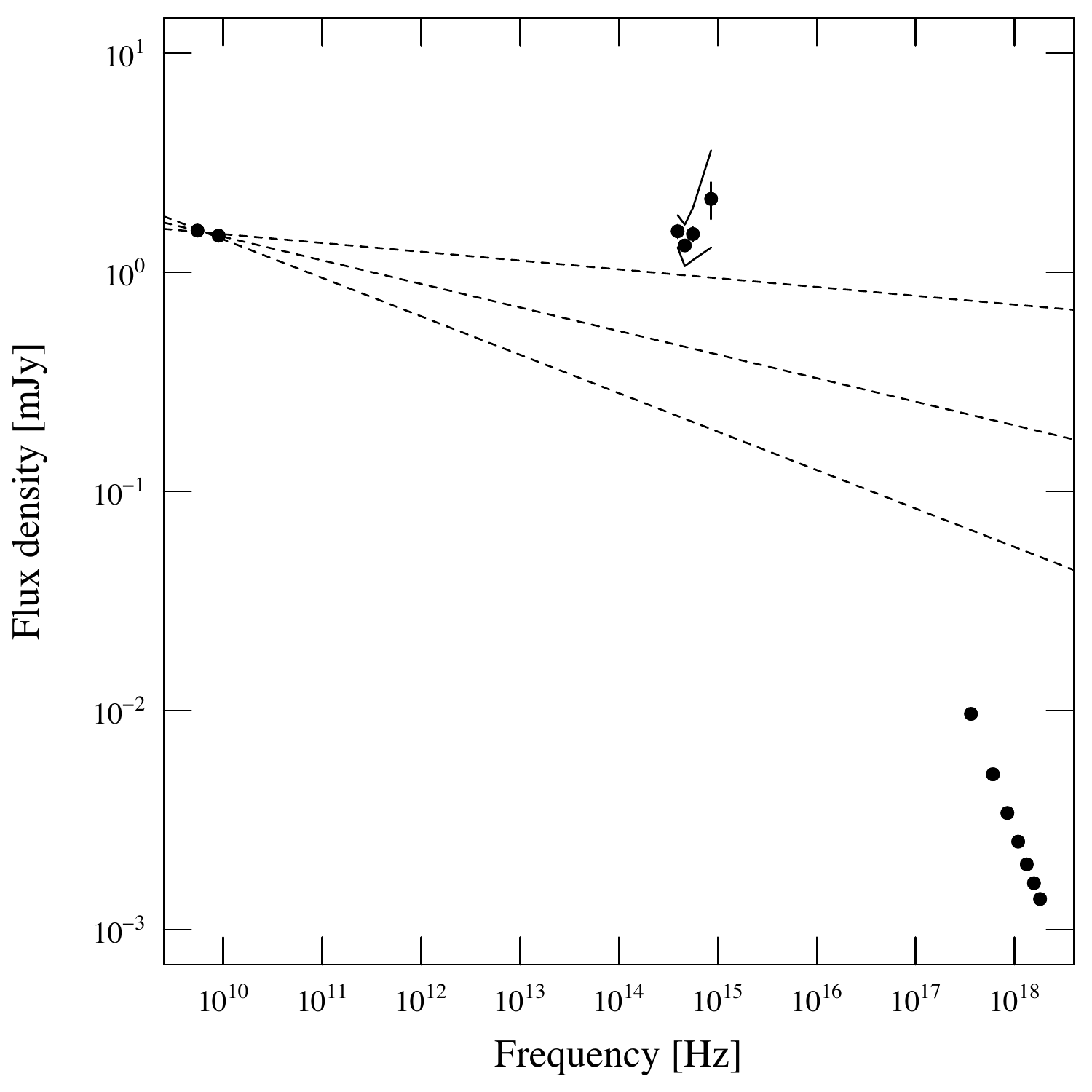}
\end{center}
\vspace{-12pt}
\caption{The multiwavelength SED of GS 1354--64. The dotted lines show the possible slopes of the power law spectrum from synchrotron radiation based on the 1$\sigma$ errors of the radio data points. The solid lines in the optical/UV frequencies show the range of the SED with varying values for A$_{V}$ (2.6$\pm$0.31). The \swiftxrt\/ X-ray spectrum has a low countrate and thus it has been binned to seven bands from 1.5 to 7.5 keV for clarity.} \label{sed}
\end{figure}

The multiwavelength SED from radio to X-rays (around MJD 57190) is plotted in Fig. \ref{sed}. The observations are not strictly simultaneous: the radio were observed on MJD 57189.32, UV/X-ray on MJD 57190.53, and we took an average of the optical magnitudes  observed on MJD 57188.99 and MJD 57190.02 for $R$-band, and on MJD 57187.01 and MJD 57193.69 for $i'$- and $V$-bands. The radio spectrum is approximately flat with spectral index $\alpha = -0.10\pm0.05$ ($S = \nu^{\alpha}$; \citealt{coriat15}). Similar values were also reported in \citep{brocksopp01} for the 1997 outburst: flat or slightly inverted radio spectra were observed during the outburst peak and decay. They could extend the flat radio spectrum to the IR and up to the $R$-band. However, the fluxes in $B$- and $V$-bands were well above the extended flat radio spectrum and likely corresponded to a thermal disc spectrum. In Fig. \ref{sed}, we show the possible slopes of a power law spectrum fitted to the radio data. It seems possible to extend the flat spectrum to optical wavelengths without a cutoff. The $U$-band flux is above the extended radio spectrum by $\sim2\sigma$, and appear bluer ($\alpha>0$), thus likely corresponding to thermal disc spectrum. Overall, the extended radio spectrum does not seem to fully explain the total optical fluxes. 

\subsection{Long term optical monitoring} \label{longmon}

\begin{figure}
\begin{center}
\includegraphics[width=0.5\textwidth]{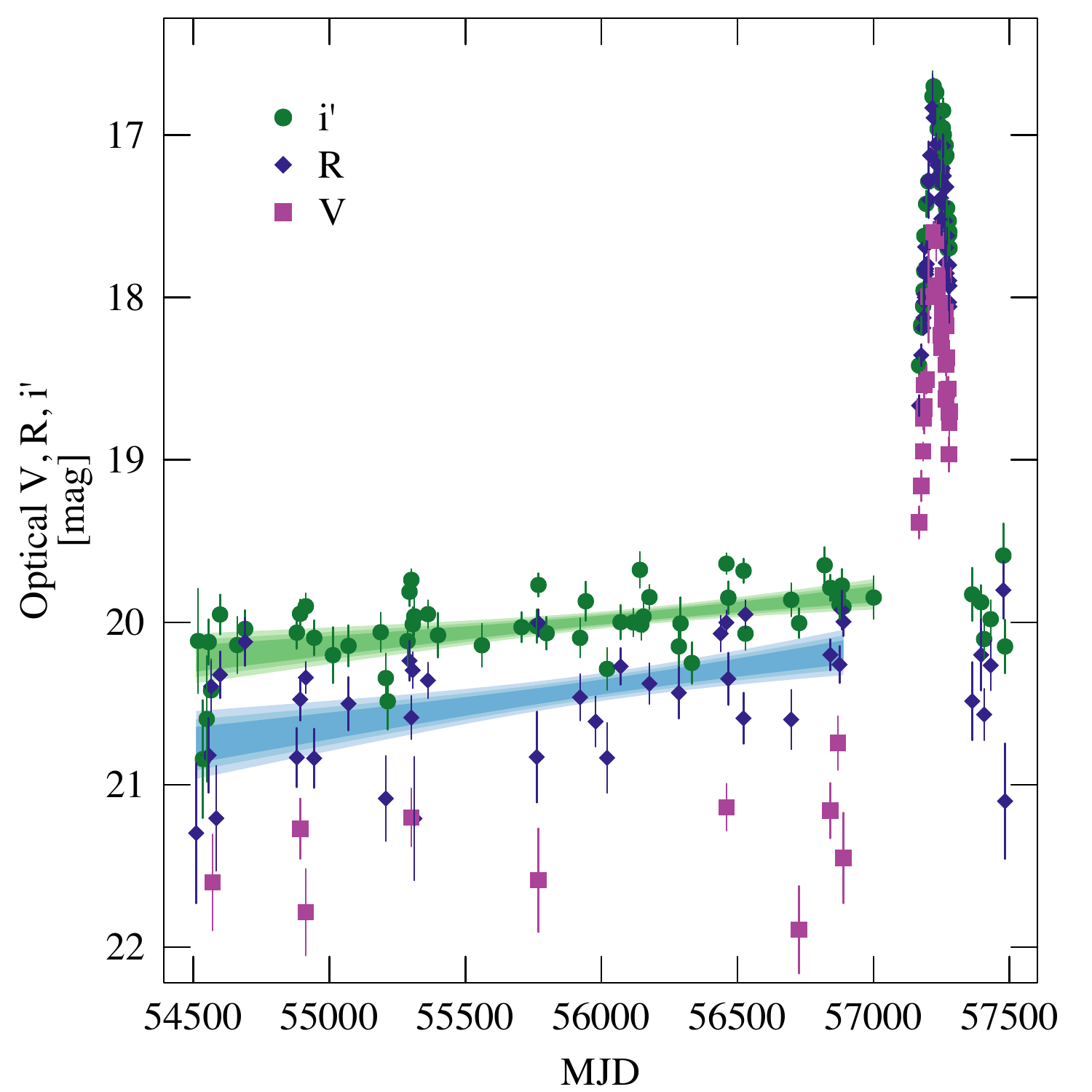}
\end{center}
\vspace{-12pt}
\caption{Long term lightcurves of the optical bands. The shaded, coloured regions show the 95\%, 99\% and 99.9\% confidence intervals on the linear regression. The colour scheme is the same as in Fig. 1.} \label{longlight}
\end{figure}

In Fig. \ref{longlight} we show the complete lightcurves from optical $i'$-, $R$- and $V$-bands that include the outburst as well as a part of the quiescent state ($\sim$ 7 years), and a few observations after the outburst. During quiescence the optical magnitudes vary by a magnitude. The orbital variability is approximately 0.1 mag \citep{casares09}, so most of the variability can be attributed to aperiodic flaring activity. We can also identify a statistically significant slow rise in the $i'$- and $R$-band (not enough data points for the $V$-band). We calculate linear least-squares regressions using Monte Carlo bootstrap methods \citep{curran14} from samples that are randomly selected and perturbed according to the data errors. We note the slope and intercept of each fit, and produce 95\%, 99\% and 99.9\% limits on the possible regressions (shaded, coloured regions in Fig. \ref{longlight}). We also note the percentage of cases where the slope is 0 or negative, and calculate the significance of a positive slope, which in both cases is $>5\sigma$. 

\begin{figure}
\begin{center}
\includegraphics[width=0.5\textwidth]{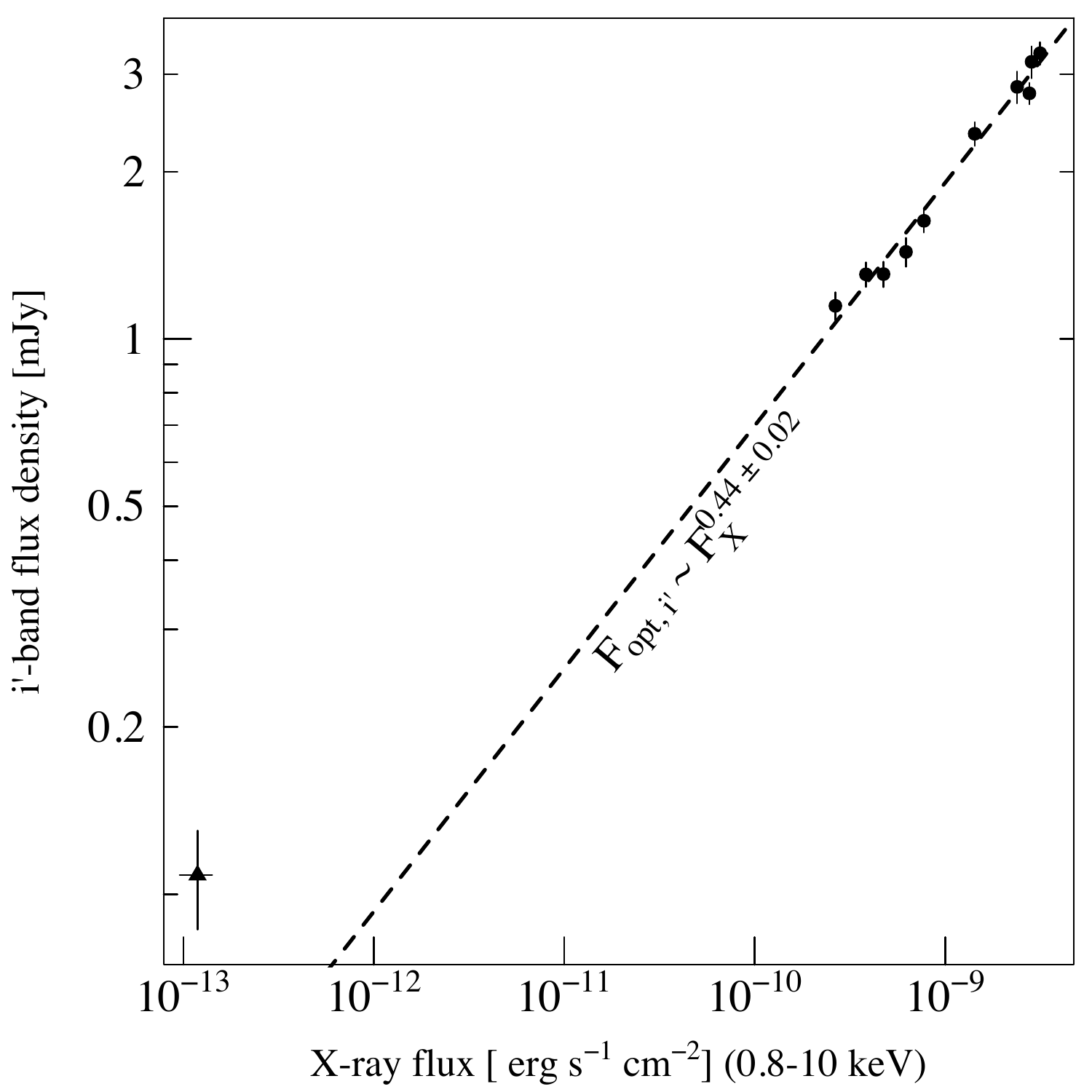}
\end{center}
\vspace{-12pt}
\caption{The optical i'-band flux density as a function of 0.8--10 keV X-ray model flux with an additional data point from the quiescent state observed with \textit{Chandra} \citep{reynolds11}. The dashed line shows the best least squares fit to the data, when excluding the quiescent data point. The quiescent data point lies well above the best-fit line, and thus it is likely that the optical/X-ray relation turns to a shallower one in the quiescent state as compared to the outburst.} \label{optixcorr}
\end{figure}

In addition, we include a quiescent data point to the optical/X-ray correlation (Fig. \ref{optixcorr}) based on \textit{Chandra} observation \citep{reynolds11} that took place on MJD 55470. There are no simultaneous optical observations during the \textit{Chandra} observation, so we estimate the optical $i'$-band magnitude using the above-mentioned linear least-squares regression to the quiescent data with an error based on the standard deviation of all quiescent data points, and assuming that the stellar contribution can be as high as is 50\% \citep{casares04,casares09}. It seems that the $F_{i'} \sim F_{X}^{0.44}$ relation does not continue to the quiescent state and a break between the outburst and quiescent state is evident ($>3\sigma$ away from the best-fit line). This suggests that the viscously heated disc with a shallower relation ($F_{opt} \sim F_{X}^{0.3}$; \citealt{russell06}) may dominate the optical emission, rather than X-ray reprocessing, at this low X-ray luminosity.

\section{Discussion} \label{discussion}

Many LMXBs show hard, or ``failed'', outbursts, where the source spectral state does not change to the soft state \citep{tetarenko15}. These sources differ in orbital periods to each other, and thus sizes of the accretion discs, which leads to different amounts of accumulated matter in the disc. In addition, these sources present very different lightcurve morphologies \citep{brocksopp04}. The 2015 outburst of GS 1453--64 seems to be a relatively ``clean'' outburst, with no secondary maxima (with the exception of optical flares), and all frequencies above optical seem to be correlated with each other. The 2015 outburst has been so far the best covered outburst of GS 1354--64 throughout the electromagnetic spectrum, which gives us another well-sampled data set for the hard outbursts detected from X-ray transients. 

\subsection{Evolution of the outburst} 

Accretion discs in LMXBs are strongly irradiated by the X-ray emission of the accretion flow. We discuss below (Section \ref{optem}) that the optical emission is likely partly produced by the reprocessing of the X-ray emission. The irradiation affects the disc properties during the outburst. According to the disc instability model (DIM; e.g. \citealt{lasota01}), during quiescence the accretion disc is replenished, increasing steadily in temperature, until some annulus becomes thermally unstable allowing for ionisation of hydrogen. The ionisation in turn increases the viscosity of the disc via a magneto-rotational instability \citep{balbus91}, which increases the mass accretion rate through the disc. This leads to an outburst, where a heating front moves inwards and/or outwards to smaller/larger radii, increasing the mass accretion rate, that increases the X-ray radiation and reprocessing, which in turn helps the heating front propagate further out. In the case of an ``inside-out'' burst, the heating fronts propagate slowly to larger radii resulting in slow rise times and symmetric outburst profiles. When the heating front arrives to the outer disc rim, there is a surface density and accretion rate excess that starts to diffuse inwards until the accretion rate is roughly constant in the whole disc. The accretion rate increases by a factor of three during this phase.

The behaviour of GS 1354--64 outburst in 1997 as well as the current outburst fits well with the above-mentioned ``inside-out'' burst. The increase in the mass accretion rate after the heating front has arrived to the outer disc could be seen in the increasing temperature of the disc, and the increasing frequency of the mHz QPO from 50 mHz to 200 mHz, corresponding to the Keplerian radii of log (r/r$_{g}$) = 4.3--4.1 (or $3\times 10^{10}$ cm -- 2$\times 10^{10}$ cm) for a black hole of 10 $M_{\odot}$, assuming that the origin of the QPOs is geometrical and related to disc truncation. This radius could correspond to the location where the disc changes from an optically thick, geometrically thin disc to an optically thin, geometrically thick disc (e.g. ADAF/RIAF; \citealt{yuan03}). In the 1997 outburst of GS 1354--64 as well as in other sources, e.g. GRO J1719--24 and XTE J1118$+$480 \citep{brocksopp04}, the evolution of the QPO is similar with the centroid frequency increasing during the peak/decay of the outburst. In the 2015 outburst the QPO frequency is about five times higher than in the 1997 outburst. At the same time the soft X-ray luminosity is also several times brighter when compared to the 1997 outburst. This is in line with the truncated disc scenario where the accretion disc approaches closer to the black hole, increasing the QPO frequency and emitting more radiation in the X-ray regime, thus providing more seed photons for the Comptonisation and subsequently increasing the X-ray brightness of the source. As the accretion disc pushes inwards into the hot accretion flow, it cools the corona by Compton scattering, which is seen as the softening of the X-ray spectra, but not enough to make the source transit to the soft state. This also happened in the 1997 outburst \citep{revnivtsev00}. Finally, the irradiation prevents the formation of a cooling front, and the disc is drained by viscous accretion of matter \citep{king98}, which produces an exponential decay of the lightcurve, which is seen in the outburst decay of GS 1354--64. It is interesting to note that the index of the exponential rise/decay changes during the outburst. This could arise from changes in the radiative efficiency as discussed in \citet{eckersall15}. They found that a factor of $\sim$ 2 change in the decay time scale is expected assuming a change from radiatively efficient emission in the soft state to radiatively inefficient emission in the hard state. In the 2015 outburst GS 1354--64 did not exhibit a state transition, but it has been suggested that some sources may change into a radiatively efficient accretion state in the hard state as well (e.g. H1743--322; \citealt{coriat11}). 
  
The Comptonised X-ray spectra show that there is little direct emission from the accretion disc. The seed photon temperature stays $<$0.4 keV, while in soft state LMXBs the temperature is usually $\sim$1 keV. Therefore, the mass accretion rate does not become large enough for the disc to condensate from the optically thin accretion flow, which dominates the X-ray emission. \citet{meyer04} consider L$_{X}$/L$_{Edd} < 0.05$ as a requirement for the evaporation/condensation scheme to hold. Observationally, this limit can be as high as L$_{X}$/L$_{Edd} < 0.11$ \citep{tetarenko15}. For larger luminosities the disc should approach the innermost stable orbit. 

\subsection{A ``high-hard'' outburst}

In Table \ref{models} the average outburst peak unabsorbed 0.8--100 keV luminosity is $1.03 \times 10^{-8}$ erg s$^{-1}$ cm$^{-2}$. In Eddington units this value can be presented as L$_{X}$/L$_{Edd}$ = 0.01 d[kpc]$^2$ / M$_{BH}$[M$_{\odot}$]. For values of d = 25 kpc and M$_{BH}$=10 M$_{\odot}$, the Eddington ratio is L$_{X}$/L$_{Edd}$ = 0.6. A similar flux level was reached in the 1987 outburst, but this time the source was reported to have a soft state X-ray spectrum (the luminosity reported was $3.5 \times 10^{37}$($d$/10 kpc) erg s$^{-1}$ in the energy band 1--10 keV; \citealt{kitamoto90}). Therefore, the current outburst was likely to have been on the brink of turning to a soft state. However, the critical luminosity for state transition is variable between individual sources, and even between individual outbursts of a same source (GX 339--4; e.g. \citealt{belloni06}). \citet{tetarenko15} found that the limiting Eddington ratio for hard-only outburst is L$_{Edd} < 0.11$. However, the peak luminosity of the 2015 outburst is several times brighter without the source transiting to the soft state. This might either indicate that this limit is higher, or that the distance to GS 1354--64 is over-estimated. 

Extrapolating and integrating over the bolometric X-ray flux gives a rough estimate of the X-ray fluence of the outburst: 0.04 erg cm$^{-2}$, or $3 \times 10^{45}$ erg for a distance of 25 kpc. With the standard efficiency of $\epsilon \sim$ 0.1, this corresponds to $3 \times 10^{25}$ g of matter accreted. If the accretion disc was emptied during both the previous and current outbursts, we can estimate the average mass accretion rate as 5$\times 10^{16}$ g/s (or 8.5$\times 10^{-10} M_{\odot}$/yr). 

We can estimate an upper limit for the mass of the black hole given the mass ratio $q=0.13\pm0.02$, and that the type of the companion star is G0 III--G5 III \citep{casares04}. A normal G0 III--G5 III star has typically mass in the range 2.1--2.4 M$_{\odot}$. However, due to mass loss, the companion is likely a lower mass and more evolved star. \citet{casares04} estimate that the companion mass should be less than 2.1 M$_{\odot}$ based on the orbital period and the resulting estimate of the Roche Lobe equivalent stellar radius, spectral classification and using the Stefan-Boltzmann relation. This results in the upper limit for the black hole mass of 16$\pm$3 M$_{\odot}$, and subsequently for the Eddington luminosity of L$_{X}$/L$_{Edd} > $ 0.38$^{+0.09}_{-0.06}$ for a distance of 25 kpc. Even for the minimum distance (15.6 kpc) as suggested by \citet{reynolds11}, the Eddington luminosity is greater than 0.15$^{+0.04}_{-0.02}$.   

A high Eddington luminosity is unusual for the canonical hard X-ray state. We rule out that GS 1354--64 would have been in a hard-intermediate state (HIMS) by its X-ray spectral and timing properties. The X-ray spectra is best-fitted with a thermal Comptonisation model, i.e. approximately a cut-off powerlaw ($\Gamma = 1.2-1.7$), while in the HIMS the power law is steeper ($\Gamma>$ 2.1) and usually without any cutoff. As the accretion disc is not statistically required in the fits to the \swift\/ spectra, the accretion disc fraction has to be small and under values observed in the HIMS where the disc usually contributes to the spectra more than 20\%. The observed QPO frequencies are also very low suggesting a hard state, while generally the QPO frequencies in the HIMS are higher ($>$1 Hz; \citealt{vignarca03}). Possible comparable sources to GS 1354--64 showing a luminous hard state are GRS 1915$+$105 and V404 Cyg. Since its discovery, GRS 1915$+$105 has been constantly found accreting at a high luminosity, including a state what could be categorised as a hard state ($L_{X}/L_{Edd} \sim$ 0.1--0.4; \citealt{peris15}). However, due to the accretion disc being very hot in this source ($>$1 keV), the disc fraction is very high unless $L_{X}/L_{Edd} \sim$ 0.1. In addition, the X-ray timing properties of GRS 1915$+$105 resemble more HIMS than hard state \citep{reig03}, and thus it is not clear if GRS 1915$+$105 displays the canonical hard X-ray state. V404 Cyg can produce hard outbursts that reach $L_{Edd}$ \citep[e.g.][]{zycki99,rodriguez15}. Especially in the 1989 outburst, the spectral model is similar to the one fitted in this paper with low electron temperature and slightly higher optical depth ($kT_{e} \sim$ 10 keV, $\tau \sim$ 6.5; \citealt{zycki99}), but the model to fit the spectra from June 2015 outburst are more complex with higher electron temperature ($kT_{e} \sim$ 40 keV) and the seed photon temperature of the Comptonization being very high ($\sim$ 7 keV; \citealt{natalucci15}). In any case, the overall outburst evolution of V404 Cyg is completely different compared to GS 1354--64, as V404 Cyg display multiple flares, shows spectral variability in minute timescales, probably because of variable absorption in the line-of-sight, and changes X-ray state during outburst and during individual flares \citep{zycki99,rodriguez15,radhika16}. The energy specta of V404 Cyg during an outburst is also markedly different from GS 1354--64, showing a strong (up to 0.6 $L_{Edd}$), un-Comptonised disc component \citep{zycki99,radhika16}. Thus, it is not clear if V404 Cyg either displays the canonical hard X-ray state with a very high luminosity.     

\subsection{Optical/UV emission processes} \label{optem}

After the outburst peak, the optical ($V$- and $R$-bands) drop more quickly compared to the X-ray and turn to an anti-correlation, reminiscent of the behaviour of other black hole transients when they switch to the soft X-ray state, e.g. XTE J1550--564 \citep{russell07} and GX 339--4 \citep{coriat09}. In both cases, this was found to be associated with a change in optical colour, and was interpreted as due to the jet component increasing in the optical/IR after the soft-to-hard state transition. However, this is not the case in GS 1354--64, as the source is in the hard X-ray state during the whole outburst, and we do not have evidence for a change in the optical colour. After the optical drop, the relation turns to an anti-correlation because of periods of brightening in what seems to be a series of flares in the optical lightcurve. A counterpart to these flares is not seen in the X-ray lightcurves, and thus we can rule out the flares occurring in the optically thin accretion flow. Optical reflares have been suggested to arise from a sequence of heating and cooling front reflections in the accretion disc \citep{menou00}. If the optical reflare is produced in the accretion disc, it should produce a different correlation coefficient with the X-rays (0.3 as discussed in Section \ref{longmon}). Assuming the quiescent state emission is dominated by the disc this correlation should span the quiescent state value as well, which is the case as found in Section \ref{longmon}. However, the disc irradiation should inhibit the formation of cooling fronts \citep{dubus01}, thus presenting doubts for this mechanism. Another possible mechanism could be flares in the jet, that would increase the synchrotron flux in the optical bands. From Fig. \ref{optsed} (right panel) we see that all the bands rise and decay in unison, thus the jet contribution should be approximately equal in all bands, i.e. more or less a flat spectrum (consistent with the radio spectral index $-0.10\pm0.05$). The multiwavelength SED in Fig. \ref{sed} demonstrates that if the jet spectrum is continued to optical frequencies without a cutoff, it can present a sizeable contribution in the optical bands. Due to the scatter and larger error bars in the UV lightcurve, it is difficult to say whether the reflares are visible in the $U$-band or not. However, without radio/IR coverage over the outburst the jet contribution is difficult to ascertain.     

In Section \ref{optx} the optical, UV and X-ray lightcurves were found to be correlated with $L_{opt/UV} \approx L_{X}^{0.4-0.5}$. This value is consistent with the X-ray irradiated accretion disc \citep{vanparadijs94} with a possible contribution from the viscous disc \citep{russell06}. Based on one multiwavelength SED (Fig. \ref{sed}) the jet contribution cannot explain (if at all) the whole optical flux, thus it is likely that the reprocessing contributes in the optical/UV frequencies. A similar value ($L_{opt} \approx L_{X}^{0.56}$) was obtained for V404 Cyg in the hard state during an outburst, which was resolved to be a combination of the contributions from the jet and reprocessing (Bernardini et al. 2016, subm.), where the slightly higher correlation coefficient, as compared to GS 1354--64, could arise from a bigger jet contribution. In the case of GS 1354--64, it is possible that the optical/UV and their correlations with the X-ray are affected by the viscous disc. For viscous disc, the correlation coefficient is expected to get smaller with longer wavelengths and the correlation coefficient of optical/X-ray should be $\sim$0.3 \citep{russell06}, which are both not observed, but the contribution from the viscous disc could lower the correlation coefficient from the theoretical value of 0.5 for irradiation. In addition, as the UV/optical SEDs are flat during the outburst, this argues against the UV/optical originating from the Rayleigh-Jeans part of a blackbody spectrum. However, the SEDs could be also inverted for higher values of extinction, and thus conclusions cannot be made based on the optical/UV SEDs. Overall, it can be assumed that the during the outburst the emission from the irradiation is dominating with a smaller contribution from the viscous disc and perhaps a contribution from the jet if the jet break lies above optical/UV. In any case, the underlying scaling factor is the mass accretion rate, which drives the evolution in all bands.      

\subsection{Quiescent data}

All versions of the DIM predict increasing quiescent fluxes (by as much as 1--2 magnitudes), which is due to increasing surface density and effective temperature as matter accumulates to the disc. On the contrary, the observations of dwarf nova outbursts have shown that the quiescent fluxes are constant or decreasing \citep{smak00}. We have shown in Fig. \ref{longlight} that a statistically significant rise is observed in the optical lightcurves. Assuming that the increase (0.56 mag in 2500 days in the $R$-band) has been constant from the previous outburst ($\sim$ 6500 days), the total optical brightening would have been $\sim$ 1.5 mag, consistent with the prediction of the DIM. According to the few R-band observations taken during a ten year period after the 1997 outburst as presented in \citet[][their Fig. 4]{casares09}, the R-band magnitude went down to 21.5 supporting the trend. However, this observation was taken about four years after the outburst, and two years after this observations the source displayed a much higher magnitude reaching up to 19.5, perhaps presenting an optical flare rising above the usual quiescence level. Thus, on top of the slow optical brightening and aperiodic, order of a magnitude flaring, there are evidence of brighter flares during quiescence.       

Recently, a similar linear, optical brightening ($\Delta$V $\sim$ 0.8 mag) has been observed from Nova Muscae 1991 \citep{wu16}. Why is this optical rise seen in GS 1354--64 and Nova Muscae 1991 but not in other sources? One possibility is that in GS 1354--64 there is evidence of the accretion flow dominating the optical flux even in quiescence from its large amplitude variability \citep{casares09}. This is not the case for sources like GRO J1655--40, where the star dominates the optical emission in quiescence \citep{greene01}. In many LMXBs, the star and the accretion flow can both be detected in quiescence: e.g. XTE J1118$+$480 \citep{gelino06}, A0620--00 \citep{cantrell08} and V404 Cyg \citep{zurita04}. These LMXBs that have the accretion flow contributing or dominating to the optical emission in quiescence also show occasionally high amplitude variability during the quiescence similar to GS 1354--64 \citep{zurita03,zurita04,hynes09,yang12}. Interestingly, in V404 Cyg, a several years long slow optical fade was followed by a low amplitude, relatively fast rise of a few months prior to the June 2015 outburst when the contribution of the star was removed \citep{bernardini16}. Whereas in some systems the accretion flow dominates with little or no evidence of the star making a contribution: e.g. GX 339--4 \citep{shahbaz01} and Swift J1357.2--0933 \citep{shahbaz13,mata15}, but no long-term quiescent light curves exist for these sources. This may be the first time a slow years-timescale optical rise has been seen from an X-ray binary together with an outburst. To determine, whether this is more ubiquitous phenomena in XRBs, more long-term, continuous observations of LMXBs are needed.

GS 1354--64 is very X-ray bright in the quiescence ($L_{X} \gtrsim 10^{34}$ erg s$^{-1} \sim 10^{-5} L_{\mathrm{Edd}}$; \citealt{reynolds11}) as compared to other LMXBs ($\sim 10^{31}$ erg s$^{-1}$). This implies a mass accretion rate $\dot{M} = L_{X}/ \epsilon c^{2} \sim \epsilon^{-1} 10^{13}$ g/s. The efficiency, $\epsilon$, varies in ADAF solutions depending on the mass accretion rate and the efficiency of electron heating, but generally lies in the region 0.001--0.05 \citep{yuan14}. Thus, the mass accretion rate in the quiescence state is $2\times10^{14}$ g/s to $10^{16}$ g/s (or $10^{-4}$ to 7$\times 10^{-3}$ in Eddington accretion rates for a 10$M_{\odot}$ black hole). These values are below the average mass accretion rate as obtained from the outburst energetics, and thus reasonable in the context of DIM. The range of values for the efficiency allow also accretion rates values that are less than the critical value of ADAF solutions ($\dot{M}_{crit, ADAF} \sim 10^{-3}$ for $\epsilon > 0.007$), and correspond to a scenario where the electrons radiate efficiently, but where the Coulomb collisions are still inefficient and thus the ions transfer only a small fraction of their energy to the electrons and remain advection dominated \citep{yuan14}. Thus, the quiescent data is consistent with the DIM including disc evaporation, where the inner accretion flow is of a bright ADAF type.   

\section{Conclusions} \label{conclusions}

We have studied in detail the evolution of the 2015 outburst of GS 1354--64 in the optical, UV and X-ray frequencies. The outburst was found to stay in the hard X-ray state despite reaching $\sim$60\% of the Eddington luminosity (for a distance of 25 kpc and M$_{BH} = 10 M_{\odot}$), thus presenting us one of the best example of a hard, or ``failed'', outburst with multiwavelength coverage throughout the outburst. We found that the optical emission, as well as the UV emission, is tightly correlated with the X-ray emission with a correlation coefficient of $\sim$0.4--0.5, that is consistent with irradiation contributing to the optical/UV frequencies, with a smaller contribution from the viscous disc. The X-ray spectra can be fitted well with a Comptonisation model, and show softening towards the end of the outburst. In addition, we detect a QPO in the X-ray lightcurves with increasing centroid frequency during the peak and decay parts of the outburst. As discussed above, this behaviour could be explained by the DIM with irradiation and disc evaporation/condensation. Finally, we studied the long-term optical lightcurves and found a statistically significant, slow rise of the source brightness at optical frequencies during the quiescence state. We argue that this behaviour is the much-sought observational evidence of matter slowly accumulating in the accretion disc, and subsequently getting optically brighter, as predicted by the DIM. The reason why this is not observed in other sources is proposed to be the unusually bright accretion disc dominating the optical emission in the quiescent state of GS 1354--64.   

\section*{Acknowledgements}

We acknowledge financial support from CONICYT-Chile grants FONDECYT Postdoctoral Fellowship 3140310 (JMC-S), FONDECYT 1141218 (FEB), Basal-CATA PFB-06/2007 (JMS-C, FEB), ``EMBIGGEN'' Anillo ACT1101 (FEB). The SMARTS data is from proposals CN2015B--81, CN2015B--88, CN2014B--044 and DD-15A-0001. The Faulkes Telescopes are maintained and operated by Las Cumbres Observatory Global Telescope Network.  ANDICAM@SMARTS 1.3m is operated by the SMARTS Consortium. This research has made use of data, software and/or web tools obtained from the High Energy Astrophysics Science Archive Research Center (HEASARC), a service of the Astrophysics Science Division at NASA/GSFC and of the Smithsonian Astrophysical Observatory's High Energy Astrophysics Division. This research made use of MAXI data provided by RIKEN, JAXA and the MAXI team, and \swiftbat\/ transient monitor results provided by the \swiftbat\/ team.

\bibliographystyle{mnras}

\bibliography{references}

\appendix
\section{Optical monitoring data from Faulkes/LCOGT and SMARTS}

\begin{table*} \centering
\caption{Optical monitoring data from Faulkes/LCOGT $i'$-band. FTS = Faulkes Telescope South; L[X] = LCOGT, with [X] denoting the telescope number.} 
\label{fts_i}
\begin{tabular}{cccc|cccc|cccc} \hline
MJD & i' & err & Tel & MJD & i' & err & Tel & MJD & i' & err & Tel \\
& mag & mag & & & mag & mag & & & mag & mag & \\\hline
54517.623 & 20.11 & 0.32 & FTS & 56070.377 & 20.00 & 0.11 & FTS & 57221.137 & 16.70 & 0.05  & L5  \\
54535.433 & 20.84 & 0.36 & FTS & 56117.367 & 20.00 & 0.09 & FTS & 57230.452 & 16.74 & 0.07  & L11 \\
54549.427 & 20.59 & 0.39 & FTS & 56141.454 & 19.68 & 0.11 & FTS & 57234.390 & 16.88 & 0.05  & L11 \\
54556.436 & 20.12 & 0.14 & FTS & 56146.501 & 20.01 & 0.09 & FTS & 57235.740 & 16.96 & 0.07 & L12   \\
54565.421 & 20.42 & 0.19 & FTS & 56154.532 & 19.96 & 0.10 & FTS & 57240.826 & 16.85 & 0.07 & L13   \\
54598.394 & 19.95 & 0.12 & FTS & 56176.388 & 19.84 & 0.08 & FTS & 57248.412 & 17.30 & 0.07 & L11   \\
54661.600 & 20.14 & 0.18 & FTS & 56284.648 & 20.15 & 0.16 & FTS & 57249.024 & 17.15 & 0.07 & L5   \\
54690.430 & 20.04 & 0.12 & FTS & 56290.699 & 20.01 & 0.16 & FTS & 57249.733 & 17.23 & 0.06 & L12   \\
54880.673 & 20.06 & 0.10 & FTS & 56332.503 & 20.25 & 0.13 & FTS & 57250.732 & 17.21 & 0.11 & L12   \\ 
54892.667 & 19.95 & 0.09 & FTS & 56459.467 & 19.64 & 0.06 & FTS & 57253.441 & 17.10 & 0.04 & L11   \\
54913.686 & 19.90 & 0.08 & FTS & 56466.392 & 19.85 & 0.10 & FTS & 57253.744 & 17.06 & 0.07 & L13   \\
54944.396 & 20.10 & 0.11 & FTS & 56522.417 & 19.68 & 0.07 & FTS & 57254.752 & 16.96 & 0.07  & L12 \\
55013.387 & 20.20 & 0.17 & FTS & 56529.489 & 20.07 & 0.10 & FTS & 57255.719 & 16.85 & 0.08 & L10   \\
55070.466 & 20.14 & 0.13 & FTS & 56697.502 & 19.86 & 0.10 & FTS & 57258.393 & 17.00 & 0.07 & L3   \\
55188.659 & 20.06 & 0.12 & FTS & 56726.708 & 20.00 & 0.09 & FTS & 57263.032 & 17.14 & 0.23 & L5   \\
55207.594 & 20.34 & 0.15 & FTS & 56820.343 & 19.65 & 0.11 & FTS & 57265.016 & 17.06 & 0.05  & L5 \\
55214.632 & 20.49 & 0.17 & FTS & 56841.361 & 19.79 & 0.08 & FTS & 57266.396 & 17.60 & 0.05 & L11  \\
55287.445 & 20.12 & 0.15 & FTS & 56869.517 & 19.84 & 0.10 & FTS & 57267.015 & 17.45 & 0.05 & L5  \\
55294.635 & 19.81 & 0.09 & FTS & 56875.367 & 19.89 & 0.09 & FTS & 57267.762 & 17.13 & 0.06 & L12  \\
55301.443 & 19.74 & 0.07 & FTS & 56883.501 & 19.77 & 0.10 & FTS & 57268.762 & 17.59 & 0.06 & L10  \\
55306.563 & 20.01 & 0.11 & FTS & 56889.449 & 19.90 & 0.10 & FTS & 57270.395 & 17.45 & 0.06  & L11 \\
55312.537 & 19.96 & 0.12 & FTS & 57000.721 & 19.85 & 0.13 & FTS & 57272.375 & 17.70 & 0.08 & L11  \\
55362.388 & 19.95 & 0.09 & FTS & 57167.540 & 18.42 & 0.05 & FTS & 57275.376 & 17.69 & 0.06 & L11  \\
55399.463 & 20.08 & 0.14 & FTS & 57175.581 & 18.17 & 0.05 & FTS & 57276.377 & 17.53 & 0.05 & L11  \\
55560.707 & 20.14 & 0.13 & FTS & 57175.590 & 18.18 & 0.05 & FTS & 57277.992 & 17.61 & 0.06 & L5  \\
55706.528 & 20.03 & 0.09 & FTS & 57182.453 & 18.05 & 0.03 & FTS & 57278.380 & 17.59 & 0.05 & L11  \\
55762.367 & 20.02 & 0.10 & FTS & 57184.001 & 17.96 & 0.06 & L5 & 57279.376 & 17.70 & 0.05 & L11 \\
55768.534 & 19.77 & 0.07 & FTS & 57186.522 & 17.84 & 0.06 & FTS & 57362.736 & 19.83 & 0.16 & FTS  \\
55798.460 & 20.07 & 0.10 & FTS & 57187.005 & 17.62 & 0.07 & L5 & 57395.640 & 19.88 & 0.11 & FTS  \\
55921.677 & 20.10 & 0.12 & FTS & 57193.693 & 17.42 & 0.08 & L13 & 57406.731 & 20.10 & 0.12 & FTS  \\
55942.583 & 19.87 & 0.12 & FTS & 57201.713 & 17.29 & 0.13 & L13 & 57430.744 & 19.98 & 0.12 & FTS  \\ 
56020.669 & 20.29 & 0.13 & FTS & 57216.699 & 16.76 & 0.16 & L13 & 57477.286 & 19.59 & 0.20 & L5 \\ 
\hline
\end{tabular}
\end{table*}

\begin{table*} \centering
\caption{Optical monitoring data from Faulkes/LCOGT $R$-band. FTS = Faulkes Telescope South; L[X] = LCOGT, with [X] denoting the telescope number.} 
\label{fts_R}
\begin{tabular}{cccc|cccc|cccc} \hline
MJD & i' & err & Tel & MJD & i' & err & Tel & MJD & i' & err & Tel \\
& mag & mag & & & mag & mag & & & mag & mag & \\\hline
54510.546 & 21.30 & 0.43 & FTS & 56459.472 & 20.00 & 0.07 & FTS & 57253.368 & 17.42 & 0.03 & L11 \\
54556.440 & 20.82 & 0.23 & FTS & 56466.397 & 20.35 & 0.16 & FTS & 57253.741 & 17.31 & 0.07 & L13 \\
54565.424 & 20.39 & 0.15 & FTS & 56522.422 & 20.59 & 0.16 & FTS & 57254.749 & 17.20 & 0.05 & L12 \\
54584.571 & 21.21 & 0.32 & FTS & 56529.494 & 19.95 & 0.09 & FTS & 57255.717 & 17.07 & 0.07 & L10 \\
54598.397 & 20.32 & 0.14 & FTS & 56697.507 & 20.60 & 0.18 & FTS & 57258.390 & 17.25 & 0.07 & L3 \\
54690.433 & 20.12 & 0.15 & FTS & 56841.358 & 20.20 & 0.10 & FTS & 57263.029 & 17.53 & 0.28 & L5 \\
54880.676 & 20.83 & 0.18 & FTS & 56875.364 & 20.26 & 0.11 & FTS & 57265.014 & 17.32 & 0.05 & L5 \\
54892.670 & 20.48 & 0.13 & FTS & 56883.498 & 19.92 & 0.12 & FTS & 57266.394 & 17.78 & 0.04 & L11 \\
54913.689 & 20.34 & 0.10 & FTS & 56889.446 & 20.00 & 0.09 & FTS & 57266.403 & 17.79 & 0.04 & L11 \\
54944.399 & 20.84 & 0.18 & FTS & 57167.537 & 18.67 & 0.06 & FTS & 57267.012 & 17.69 & 0.04 & L5 \\
55070.469 & 20.50 & 0.16 & FTS & 57175.587 & 18.36 & 0.07 & FTS & 57267.759 & 17.32 & 0.05 & L12 \\
55207.597 & 21.09 & 0.26 & FTS & 57182.450 & 18.19 & 0.04 & FTS & 57268.760 & 17.85 & 0.06 & L10 \\
55294.638 & 20.24 & 0.12 & FTS & 57184.007 & 18.12 & 0.05 & L5 & 57270.393 & 17.62 & 0.05 & L11 \\
55301.446 & 20.59 & 0.13 & FTS & 57186.525 & 18.02 & 0.07 & FTS & 57272.373 & 17.89 & 0.08 & L11 \\
55306.566 & 20.30 & 0.11 & FTS & 57186.585 & 17.83 & 0.14 & L3 & 57275.374 & 17.93 & 0.05 & L11 \\
55312.539 & 21.21 & 0.38 & FTS & 57187.010 & 17.98 & 0.05 & L5 & 57277.370 & 18.03 & 0.05 & L11 \\
55362.391 & 20.36 & 0.11 & FTS & 57193.698 & 17.86 & 0.05 & L13 & 57277.989 & 17.80 & 0.05 & L5 \\
55762.373 & 20.83 & 0.28 & FTS & 57201.717 & 17.27 & 0.24 & L13 & 57278.377 & 17.90 & 0.05 & L11 \\
55768.539 & 20.00 & 0.08 & FTS & 57216.702 & 16.83 & 0.21 & L13 & 57278.732 & 18.06 & 0.10 & L12 \\
55921.682 & 20.46 & 0.14 & FTS & 57221.143 & 16.89 & 0.05 & L5 & 57279.373 & 17.93 & 0.05 & L11 \\
55978.715 & 20.61 & 0.15 & FTS & 57230.457 & 17.05 & 0.07 & L11 & 57362.733 & 20.49 & 0.24 & FTS \\
56020.674 & 20.84 & 0.22 & FTS & 57234.392 & 17.18 & 0.05 & L11 & 57395.637 & 20.20 & 0.22 & FTS \\
56070.382 & 20.27 & 0.11 & FTS & 57235.726 & 17.25 & 0.04 & L12 & 57406.728 & 20.57 & 0.16 & FTS \\ 
56176.393 & 20.38 & 0.13 & FTS & 57248.409 & 17.39 & 0.06 & L11 & 57430.741 & 20.27 & 0.15 & FTS \\
56284.654 & 20.43 & 0.16 & FTS & 57249.730 & 17.52 & 0.05 & L12 & 57477.290 & 19.80 & 0.18 & L5 \\
56438.552 & 20.07 & 0.11 & FTS & 57250.729 & 17.54 & 0.08 & L12 \\ 
\hline
\end{tabular}
\end{table*}

\begin{table*} \centering
\caption{Optical monitoring data from Faulkes/LCOGT $V$-band. FTS = Faulkes Telescope South; L[X] = LCOGT, with [X] denoting the telescope number.} 
\label{fts_V}
\begin{tabular}{cccc|cccc|cccc} \hline
MJD & i' & err & Tel & MJD & i' & err & Tel & MJD & i' & err & Tel \\
& mag & mag & & & mag & mag & & & mag & mag & \\\hline
54570.406 & 21.60 & 0.30 & FTS & 57186.582 & 18.69 & 0.15 & L3 & 57254.747 & 18.03 & 0.08 & L12 \\      
54892.672 & 21.27 & 0.19 & FTS & 57187.008 & 18.67 & 0.07 & L5 & 57255.714 & 17.86 & 0.11 & L10 \\     
54913.691 & 21.79 & 0.27 & FTS & 57193.695 & 18.51 & 0.09 & L13 & 57258.388 & 18.10 & 0.10 & L3 \\    
55301.449 & 21.20 & 0.18 & FTS & 57201.714 & 18.00 & 0.28 & L13 & 57265.011 & 18.09 & 0.08 & L5 \\      
55768.537 & 21.59 & 0.32 & FTS & 57221.140 & 17.60 & 0.07 & L5 & 57266.391 & 18.62 & 0.08 & L11 \\     
56459.470 & 21.14 & 0.15 & FTS & 57230.455 & 17.65 & 0.13 & L11 & 57266.400 & 18.63 & 0.08 & L11 \\     
56726.711 & 21.89 & 0.27 & FTS & 57234.395 & 17.93 & 0.09 & L11 & 57267.009 & 18.41 & 0.07 & L5 \\    
56841.355 & 21.16 & 0.17 & FTS & 57235.723 & 17.99 & 0.06 & L12 & 57267.757 & 18.17 & 0.08 & L12 \\     
56869.511 & 20.74 & 0.17 & FTS & 57248.406 & 18.23 & 0.08 & L11 & 57268.757 & 18.57 & 0.08 & L10 \\     
56889.443 & 21.45 & 0.28 & FTS & 57249.018 & 18.23 & 0.10 & L5 & 57270.390 & 18.37 & 0.09 & L11 \\     
57167.535 & 19.39 & 0.10 & FTS & 57249.727 & 18.25 & 0.08 & L12 & 57275.371 & 18.71 & 0.08 & L11 \\     
57175.574 & 19.16 & 0.09 & FTS & 57250.727 & 18.31 & 0.13 & L12 & 57276.372 & 18.56 & 0.08 & L11 \\    
57182.448 & 18.95 & 0.06 & FTS & 57253.366 & 18.18 & 0.06 & L11 & 57277.368 & 18.97 & 0.11 & L11 \\      
57184.004 & 18.74 & 0.07 & L5 & 57253.435 & 18.08 & 0.05 & L11 & 57278.375 & 18.77 & 0.08 & L11 \\    
57186.524 & 18.54 & 0.09 & FTS & 57253.738 & 18.11 & 0.11 & L13 &  57279.371 & 18.70 & 0.08 & L11 \\ 
\hline
\end{tabular}
\end{table*}

\begin{table} \centering
\caption{Optical monitoring data from LCOGT $B$-band. L[X] = LCOGT, with [X] denoting the telescope number.} 
\label{fts_B}
\begin{tabular}{cccc} \hline
MJD & i' & err & Tel  \\
& mag & mag & \\\hline
57221.145 & 18.64 & 0.07 & L5 \\
57234.397 & 18.97 & 0.17 & L11 \\
57235.728 & 18.83 & 0.09 & L12 \\
57240.823 & 19.06 & 0.15 & L13 \\
\hline
\end{tabular}
\end{table}

\begin{table} \centering
\caption{Optical monitoring data from SMARTS $R$-band.} 
\label{smarts}
\begin{tabular}{ccc} \hline
MJD & i' & err  \\
& mag & mag \\\hline
57188.995 & 18.00 & 0.01 \\  
57190.020 & 17.69 & 0.01 \\  
57192.004 & 17.84 & 0.01 \\ 
57193.991 & 17.82 & 0.01 \\  
57196.123 & 17.79 & 0.05 \\   
57203.081 & 17.29 & 0.01 \\  
57204.972 & 17.40 & 0.01 \\  
57208.101 & 17.12 & 0.01 \\  
57211.094 & 17.13 & 0.01 \\   
57253.006 & 17.25 & 0.01 \\ 
\hline
\end{tabular}
\end{table}

\label{lastpage}

\end{document}